\documentclass[journal]{IEEEtran}
\usepackage{amsmath,amssymb,epsfig}
\usepackage{multirow}
\usepackage{tikz, color, soul}
\usepackage{comment}
\usepackage{physics}
\usepackage{url}

\newcommand{\LL}{{\cal L}}

\newcommand\M{\mathbf{M}}
\renewcommand\P{\mathbf{P}}
\newcommand\X{\mathbf{X}}
\newcommand\Y{\mathbf{Y}}
\newcommand{\git}{{\tt \url{https://github.com/giu-guarino/R-PNN}}}

\begin{document}

\title{Band-wise Hyperspectral Image Pansharpening using CNN Model Propagation}

\author{Giuseppe~Guarino,~\IEEEmembership{Graduate~Student~Member,~IEEE,}
		Matteo~Ciotola,~\IEEEmembership{Graduate~Student~Member,~IEEE,}
        Gemine~Vivone,~\IEEEmembership{Senior~Member,~IEEE,}
        and~Giuseppe~Scarpa,~\IEEEmembership{Senior~Member,~IEEE}
        
\thanks{Giuseppe Guarino and Matteo Ciotola are with the Department
of Electrical Engineering and Information Technology, University Federico II, 80125 Napoli, Italy
(e-mail: giuseppe.guarino2@unina.it (G.G.) and matteo.ciotola@unina.it (M.C.)).}
\thanks{Gemine Vivone is with the National Research Council, 
Institute of Methodologies for Environmental Analysis, CNR-IMAA, 85050 Tito, Italy, and also
with the National Biodiversity Future Center (NBFC), 90133 Palermo, Italy
(e-mail: gemine.vivone@imaa.cnr.it).}
\thanks{Giuseppe Scarpa is with the Department of Engineering, University Parthenope, 80143 Napoli, Italy 
(e-mail: giuseppe.scarpa@uniparthenope.it).}
}

\markboth{}{}

\maketitle

\begin{abstract}
Hyperspectral pansharpening is receiving a growing interest since the last few years
as testified by a large number of research papers and challenges.
It consists in a pixel-level fusion between a lower-resolution hyperspectral datacube
and a higher-resolution single-band image, the panchromatic image,
with the goal of providing a hyperspectral datacube at panchromatic resolution.
Thanks to their powerful representational capabilities, 
deep learning models have succeeded to provide unprecedented results 
on many general purpose image processing tasks.
However, when moving to domain specific problems, as in this case,
the advantages with respect to traditional model-based approaches
are much lesser clear-cut due to several contextual reasons. 
Scarcity of training data, lack of ground-truth, data shape variability, are some such factors 
that limit the generalization capacity of the state-of-the-art deep learning networks for hyperspectral pansharpening.
To cope with these limitations,
in this work we propose a new deep learning method which inherits a simple single-band unsupervised pansharpening model nested in a sequential band-wise adaptive 
scheme, where each band is pansharpened refining the model tuned on the preceding one.
By doing so,
a simple model is propagated along the wavelength dimension, adaptively and flexibly,
with no need to have a fixed number of spectral bands,
and, with no need to dispose of large, expensive and labeled training datasets.
The proposed method achieves very good results on our datasets,
outperforming both traditional and deep learning reference methods.
The implementation of the proposed method can be found on \git.
\end{abstract}


\IEEEpeerreviewmaketitle

\section{Introduction}
\IEEEPARstart{E}{arth} observation from the space is a hot topic requiring the development of instruments and satellite platforms to face with increasing needs of data with a worldwide coverage. Unfortunately, remote sensing optical sensors show a trade-off among spatial and spectral resolutions, and signal-to-noise ratio (SNR). These trade-offs cannot be solved through hardware solutions capturing high spatio-spectral representations of the Earth without strongly penalizing the SNR. Thus, simultaneously acquiring more than one representation of the Earth with sensors showing different (often opposite) features is usually considered in the design of a payload for a satellite. Hence, in optical remote sensing, it is common to see devices having high spatial resolution but with limited spectral bands (e.g., panchromatic) working together with high spectral resolution but with lower spatial resolution sensors (e.g., multi/hyperspectral ones capturing tens/hundreds of spectral bands, respectively). Starting from the images acquired by these systems, researchers are developing software-based solutions to combine these data to get the best from each source of information. The most dated and widely used framework relies upon the fusion of panchromatic (PAN) and multispectral (MS) images. Pansharpening, which stands for panchromatic sharpening, refers to these approaches \cite{Vivo14a,Vivo21}. Other powerful examples of these techniques, often representing an extension of the pansharpening concept, are the hyperspectral (HS) pansharpening ones \cite{Lonc15}, involving HS data to be enhanced by PAN images, and MS and HS image sharpening methodologies \cite{Vivo23b}, which exploit MS data to provide high spatial resolution information to sharp a HS cube.  

HS images acquired by sensors onboard of satellite platforms are widely used for several tasks (see, e.g., classification and object detection) thanks to their very appealing spectral features. However, a limitation is represented by the spatial resolution of these data that is rarely finer than 30~m. Thus, cutting-edge research can be found in the literature fusing HS with PAN images to improve it (i.e., the so-called HS pansharpening). The relevance of this topic can be demonstrated by the recent scientific production summarized in review papers as \cite{Lonc15}, and the last HS pansharpening challenge \cite{Vivo23} organized in conjunction with the 12th WHISPERS (Workshop on Hyperspectral Image and Signal Processing: Evolution in Remote Sensing), 
which leverages on datasets from the PRISMA ({\em PRecursore IperSpettrale della Missione Applicativa}) system, 
owned and managed by the Italian Space Agency.

The first attempt of fusing real HS and PAN images captured by Hyperion/Advanced Land Imager (ALI) dated back to 2007 \cite{Capo07}, where an optimized component substitution approach inspired by the literature of multispectral pansharpening \cite{Garz08,Vivo19} has been proposed. Afterwards, many methods borrowed from the classical multispectral pansharpening literature have been tested on the HS and PAN fusion problem. A pioneering work has been presented in \cite{Vivo14a}, where several component substitution, {\em e.g.} \cite{Aiaz07, Labe00}, and multi-resolution analysis, for example \cite{Alparone2003, Aiazzi2006}, approaches designed for multispectral pansharpening have been considered. In the above-mentioned paper, an interesting study about the comparison between the fusion of HS and PAN images acquired by the same or different platforms has been proposed. 
Moreover, other interesting approaches, 
originally designed for fusion of high-resolution panchromatic and low-resolution MS data, have been adapted to the HS pansharpening case.
These can be roughly cast in Bayesian \cite{Zhang2009, Simo14, Wei2015, Wei2015a}, matrix factorization \cite{Berne2010, Moel09, Huang2013} and variational \cite{Kawakami2011} methods.
In 2015, many of these methods and others have been compared in an extensive review \cite{Lonc15}. 
Besides, several solutions specifically conceived for the HS pansharpening problem have also been proposed.
Notable examples are the use of guided filters \cite{Qu17}, variational approaches like \cite{Adde17,Huan17}, 
and the saliency analysis-based component substitution method proposed in \cite{Dong20}.

Recently, the number of research papers about sharpening of optical remote sensing data has dramatically grown, in particular related to the use of deep learning \cite{Masi2016,Yang2017, Masi2017, Deng22, Ciotola2023}. This trend is confirmed even for the HS pansharpening. 
Indeed, in 2019, a new HS pansharpening framework via spectrally predictive convolutional neural networks (CNNs) has been proposed in \cite{He19} to strengthen the spectral prediction capability of a pansharpening network. Subsequently, a dual-attention residual network upsampling the HS image using a deep HS prior module has been considered in \cite{Zhen20}. 
In \cite{He20}, a new spectral-fidelity CNN for HS pansharpening has been developed to control the spectral distortion of fused products and to progressively synthesize spatial details. 
Furthermore, a novel CNN-based method for arbitrary resolution HS pansharpening based on a two-step relay optimization process has been proposed in \cite{He22}. 
On the same research line, an arbitrary scale attention upsampling module has been introduced in \cite{He22a}.
Thus, in \cite{Band22}, an overcomplete residual network, which is focused on learning high-level features by constraining receptive fields of deep layers, has been designed together with a new spatial-domain constraint between the PAN and its predicted version. 
An unsupervised HS pansharpening method via ratio estimation and residual attention network has been described in \cite{Nie22}. 
A multistage dual-attention guided fusion network has been considered in \cite{Guan22} employing a three-stream structure and fusing the extracted features through a dual-attention guided fusion block. 
In \cite{Wu22}, it is proposed a deconvolution long short-term memories network with bi-direction learning for HS upsampling and spatial-spectral reconstruction based on a two-branch divide-and-conquer network.
A generative super-resolution network combined with a segmentation-based injection gain estimation \cite{Sale00,Rest16a} is instead proposed in \cite{Dong22}.
Finally, a deep CNN exploiting Gaussian–Laplacian pyramids for pansharpening has been presented in \cite{Dong22a}. Following the same idea of multi-resolution fusion, in \cite{Qu22a}, a multi-resolution spatial-spectral feature learning has been proposed transforming the existing deep (and complex) network into several simple and shallow subnetworks to simplify the learning process and using multi-resolution 3-D convolutional autoencoder networks to learn spatial-spectral HS features.

The 2022 WHISPERS contest on HS pansharpening was held with the goal of providing a picture of the state-of-the-art on the topic, also in light of the recent advances on deep learning, to pave the way for better solutions.
Unfortunately, none of the competitors achieved convincing results compared to the baseline methods and no winners were declared by the organizing committee.
Actually, a careful inspection of the outcomes reveals that a critical bottleneck was the limited capacity of the proposed solutions to generalize moving from synthetic reduced-resolution datasets (ground-truth available),
to real full-resolution ones (ground-truth unavailable). 
Indeed, this is one of the main problems already encountered in the case of multispectral image pansharpening using deep learning,
motivating the development of unsupervised training solutions \cite{Luo2020, Seo2020, Ma2020, Ciot2022, Ciot2023}.
In fact, unsupervised learning procedures do not require ground-truths, 
with no need to do synthetic (downgrading) resolution shifts on data.
An attempt to follow this same path for the HS case can be found in \cite{Nie22}.
However,
the wide variability of observed images, due to diversity of sensors, scenes, operating conditions, 
still prevents from generalizing well to data not seen during the training. In computer vision, this problem is usually solved by increasing the training set and by using suitable forms of augmentation \cite{Chen23}. Such solutions are hardly viable in remote sensing using hyperspectral images, due to the scarcity of high-quality training data (often proprietary) and the peculiarities of hyperspectral imaging, including the data volume per ground surface unit. 
Besides, 
compared to the multispectral case, in the HS case the resolution ratio is typically higher and the spectral coverage is much denser and
wider, exceeding the PAN bandwidth by far, causing further ill-posedness issues.
Furthermore, the number of spectral bands is a specific feature of the HS sensor
and can even change from one image to another of the same sensor, 
because of acquisition errors that can make useless subsets of bands.
A solution for handling a variable number of bands \cite{Chen21}, based on a single pretrained model, has been proposed in \cite{Qu22}.

To cope with the above issues,
in this work, we propose a new HS CNN-based pansharpening method, 
which regards the HS datacube as a chain of individual bands to be sequentially pansharpened.
This is achieved by leveraging on a lightweight single-band pansharpening network
operating in adaptive mode, whose optimized parameters for a given band are used as stating point for the self-adaptive inference step
of the next spectral band.
By doing so, 
we bridge the model parameters for adjacent bands to some extent, 
simplifying the adaptation task thanks to their expected correlation.
Both {(pre)training} and tuning iterations for target adaptation are run at full resolution thanks to a suitably defined unsupervised loss
comprising both spectral and spatial consistency terms.
It is worth to observe that the proposed solution is not just a divide-and-conquer solution 
based on the split of the HS datacube in batches of bands. 
In fact, the tuning-based protocol for bridging the models of adjacent bands 
depicts a completely new framework, 
where any baseline single-band pansharpening network, 
as well as any unsupervised loss, 
can be straightforwardly integrated.

Specifically, the advantages of the proposed solution,
hereinafter referred to as R-PNN (Rolling hyperspectral Pansharpening Neural Natwork), 
with respect to the state-of-the-art are the following.
All but the first spectral band do not need pretrained parameters, as they are inherited from the previous one.
In this way, the model progressively and adaptively learns exclusively on the target image,
with a limited computational cost thanks to several design choices such as lightweight architecture, band-wise processing, model propagation, and adaptive distribution of the tuning iterations.
Also, the method is not subject to cross-resolution generalization issues as it learns at the target resolution (being unsupervised).
More in general, the generalization is not an issue because the network learns directly on the target image, dynamically fitting its parameters to it.
Finally, the sequential structure of the method allows to handle an arbitrary number of spectral bands, not necessarily uniformly sampled.

The proposed solution has provided state-of-the-art results on all the considered datasets, both full- and (surprisingly) reduced-resolution ones, consistently outperforming all the competitors.
The above discussed properties combined with the good obtained results make the proposed method very attractive for its use in practical real-world applications.
In this perspective and to ensure full reproducibility of our research outcomes, the code of the proposed method is shared on github (\git).

In summary, the main contributions of this work are as follows:
\begin{itemize}
\item A new unsupervised CNN-based HS pansharpening approach based on a band-wise model propagation protocol.
\item A new unsupervised spectral-spatial consistency loss for PAN-HS pairs.
\item A target-adaptive solution for the PAN-HS fusion problem. 
\item State-of-the-art results on both reduced-resolution synthetic data and full-resolution real (PRISMA) data.
\end{itemize}

The reminder of the paper is organized as follows. Sec.~\ref{sec:method} presents the proposed solution. 
Sec.~\ref{sec:data} describes datasets, quality assessments indexes, and reference methods. 
Sec.~\ref{sec:validation} presents an experimental analysis aimed to support and validate several design choices.
Finally, Sec.~\ref{sec:results} gathers and discusses comparative numerical and visual results
with concluding remarks given in Sec.~\ref{sec:conclusions}. 
\section{Proposed method for HS pansharpening: R-PNN}
\label{sec:method}

Compared to the more familiar case of pansharpening of multispectral images,
in the case of hyperspectral data the generalization of deep learning models is a more critical issue for several reasons:
\begin{itemize}
\item[{a.}] lesser training datasets;
\item[{b.}] increased spectral information;
\item[{c.}] low or no correlation between the PAN and many spectral bands to be super-resolved;
\item[{d.}] variable number of bands, even for the same sensor, due to acquisition issues (different bands may be discarded for quality reasons).
\end{itemize} 
Moving from the above observations,
here we propose a CNN-based band-wise pansharpening solution
which inherits the basic idea proposed in \cite{Ciot2022} for the multispectral case,
that is to run parameters tuning iterations on the target image using a suitably defined unsupervised loss.
However, differently from \cite{Ciot2022}, 
in the present proposal, the PAN is fused with only one band at a time, rather than with all of them,
and the tuned model is passed to the next band to which is further adapted.

In the following, 
we will first detail the single-band tuning/inference block, then we will describe the overall high-level scheme,
before providing details about the core CNN network and the loss.

\subsection{Single-band pansharpening with tuning}
\begin{figure}
\centering
\includegraphics[scale=0.9]{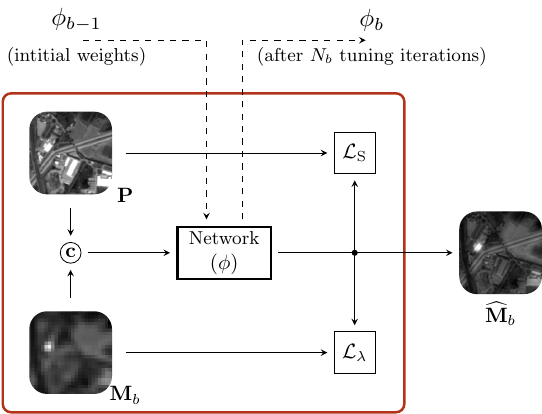}
\caption{CNN-based single-band unsupervised tuning block for pansharpening. 
The module takes in input the two images to fuse (the $b$-th band, $\M_b$, and the PAN, $\P$)
and the initial parameters, $\phi_{b-1}$, inherited from the previous module. 
It runs $N_b$ tuning iterations to fit on $\M_b$ and $\P$.
The tuned parameters, $\phi_b$, are used to provide the final pansharpened band, $\widehat{\M}_b$, and passed to the next module.
$\LL_\lambda$ and $\LL_{\rm S}$ are the spectral and spatial consistency loss terms, respectively.}
\label{fig:tuning}
\end{figure}
Fig.~\ref{fig:tuning} provides a high-level description of the tuning loop for 
the generic $b$-th spectral band, $\M_b \in \mathbb{R}^{W{\times}H{\times}1}$, to be fused with the PAN image, 
$\P \in \mathbb{R}^{RW{\times}RH}$, being $W$ and $H$ the spatial dimensions in the low resolution domain, 
$B$ the number of spectral bands, and $R$ the resolution ratio between $\P$ and the HS image $\M \in \mathbb{R}^{W{\times}H{\times}B}$.

The tuning starts from the initial network parameters $\phi_{b-1}$
fitted on $(\M_{b-1}, \P)$ in the previous step, when $b>1$, otherwise from pre-trainined weights when $b=1$.
Then, a prefixed number, $N_b$, of training iterations are run on the same target image pair, $(\M_b,\P)$,
leveraging on an unsupervised loss comprising both spectral ($\LL_\lambda$) and spatial ($\LL_{\rm S}$) consistency terms.
The tuned parameters, $\phi_b$, are then used for the final prediction of the 
pansharpened band $\widehat{\M}_b\in\mathbb{R}^{RW{\times}RH{\times}1}$,
being $\widehat{\M}\in\mathbb{R}^{RW{\times}RH{\times}B}$ the full set of HS pansharpened bands,
and passed to the next fusion step involving $\M_{b+1}$ and $\P$.
Details about the loss will be provided in Sec.~\ref{sec:unsup loss}.

\subsection{High-level model propagation scheme}
\begin{figure}
\centering
\includegraphics[scale=0.8]{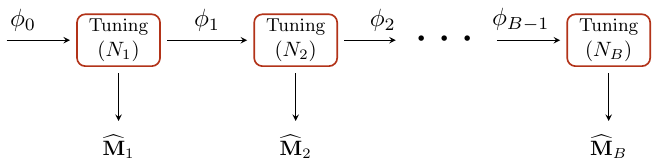}
\caption{Unsupervised rolling adaptation scheme for hyperspectral image pansharpening.
Each tuning module (detailed in Fig.~\ref{fig:tuning}) inherits the initial weights, $\phi_{b-1}$, from the previous module and passes 
the tuned ones, $\phi_b$, to the next block. $B$ is the total number of spectral bands.}
\label{fig:rolling}
\end{figure}
The overall tuning-prediction chain involving all $B$ spectral bands is depicted in Fig.~\ref{fig:rolling}.
Since the inter-band spacing can vary a lot from one pair of neighboring bands to another (there can be large gaps), 
the number of tuning iterations, $N_b$,
is increased for larger spectral gaps, $\Delta \lambda_b = \lambda_b - \lambda_{b-1}$, 
being $\lambda_b$ the $b$-th band wavelength.
This because the larger the spectral distance, the lower the expected inter-band correlation,
the lesser the fitting of the model parameters $\phi_{b-1}$ to the target band $\M_b$.
In particular, we have heuristically fixed such number as follows:
\begin{equation}
N_b = \begin{cases}
			20, & b = 1  \\
			\min\left(\alpha \Delta \lambda_b, 80 \right), & b>1
\end{cases},
\label{eq:Nb}
\end{equation}
being $\alpha$ (iter./nm) the number of iterations per nm of distance between bands $b$ and $b-1$.
As we can roughly assume a minimum spectral sampling interval of about 10nm for PRISMA, 
if $\alpha=1$, each band, $\forall b>1$, will undergo at least 10 tuning iterations up to a maximum of 80 (upper bound value).

The proposed scheme allows to drastically reduce the computational load required for parameter tuning
thanks to the stronger correlation expected between closer band pairs.
Notice that, differently form more common training/tuning configurations, where minibatches of small example patches are formed,
here, following the tuning scheme proposed in \cite{Scar2018, Ciot2022}, we have a unique training batch containing the whole target image.
Indeed, recent variants of this scheme \cite{Ciot2023, Ciot2023a} propose sampling rules to keep limited the computational burden in case for very large datasets.
In this work, however, the sizes of the interested datasets were such that no sampling was needed.

\subsection{Network}
The key characteristics of the proposed approach are its adaptivity to the target image and 
the band-wise, chained, modality. 
For these reasons it makes perfectly sense to look at lightweight network architectures
to preserve the nimbleness of the proposed solution.
Therefore, 
we decided to rely on a shallow three-layer residual model similar to the one proposed in 
\cite{Scar2018} for the classical pansharpening of 4- or 8-bands multispectral images.
It is composed of three sequential convolutional layers,
interleaved by ReLU activations,
with a parallel global skip connection 
that brings the spectral input band to be pansharpened (already interpolated to fit the PAN size)
directly to the exit (by sum) of the third convolutional layer.
In detail, 
the hyper-parameters of the network are given in Tab.~\ref{tab:hyper parameters}.
The most relevant differences compared to the CNN architecture \cite{Scar2018}
are the number of spectral bands, just one instead of 4/8, and the resolution ratio, which is 6 for PRISMA datasets.

\begin{table}
\centering
\begin{tabular}{ccccc}
\hline
layer & input bands & output bands & kernel & activation \\
\hline
1 & 2$^{(*)}$ & 48 & $7{\times}7$ & ReLU \\
2 & 48 & 32 & $5{\times}5$ & ReLU \\
3 & 32 & 1$^{(**)}$ & $3{\times}3$ & Identity \\
\hline
\end{tabular}
\vspace{1mm}
\caption{Net hyperparameters. $^{(*)}$Concatenation of $\P$ with the $b$-th (6$\times$6)${\times}$ interpolated HS band, $\widetilde{\M}_b$. 
$^{(**)}$Pansharpened band, $\widehat{\M}_b$.}
\label{tab:hyper parameters}
\end{table}

\subsection{Unsupervised spatial-spectral consistency loss}
\label{sec:unsup loss}
The proposed model leverages on a band-wise unsupervised loss, $\LL^{(b)}$, both in pre-training and fine-tuning (see Fig.~\ref{fig:tuning}), which comprises two terms,
a spectral component, $\LL_\lambda$, and a spatial component, $\LL_{\rm S}$,
weighted by a hyperparameter $\beta$:
\begin{equation}
\LL^{(b)} = \LL_\lambda\left(\widehat{\M}_b, \M_b\right)+ \beta \LL_{\rm S}\left(\widehat{\M}_b, \P\right), \;\;\; \forall b
\label{eq:L}
\end{equation}
The spectral loss term is based on a usual $\ell_1$-norm between the original low-resolution spectral band $\M_b$ and the downgraded version $\widehat{\M}_b^{(\mathcal{D})}$ of the fused band $\widehat{\M}_b$, 
obtained using an MTF-matched (Modulation Transfer Function) Gaussian Low-Pass Filtering (${\rm LPF}$) followed by decimation, 
{\em i.e.}:
\begin{equation}
 \LL_\lambda\left(\widehat{\M}_b, \M_b\right) = \left\| \widehat{\M}_b^{(\mathcal{D})} - \M_b \right\|_1,
\label{eq:LL}
\end{equation}
with
\begin{equation}
\widehat{\M}_b^{(\mathcal{D})}(n,m) \triangleq \widehat{\M}^{\rm LPF}_b(n_0+6n, m_0+6m),
\label{eq:down}
\end{equation}
being $(n_0,m_0)$ a proper spatial offset 
and $R=6$ the decimation step which equals the resolution ratio for PRISMA data.
On the other hand,
the linking of the super-resolved band, $\widehat{\M}_b$, to the panchromatic band, $\P$, 
passes through the {\em local} correlation coefficient (CC) between the two.
In particular, said $\rho^\sigma_{\X\Y}(i,j)$ the local CC between scalar images $\X$ and $\Y$, 
computed on a $\sigma{\times}\sigma$ window $w(i,j)$ centered at location $(i,j)$,
{\em i.e.},
\begin{equation}
\rho^\sigma_{\X\Y}(i,j) = \frac{{\rm Cov}\left(\X_{w(i,j)}, \Y_{w(i,j)}\right)}{\sqrt{{\rm Var}\left(\X_{w(i,j)}\right){\rm Var}\left(\Y_{w(i,j)}\right)}},
\label{eq:cc}
\end{equation}
the spatial (or structural) loss term is defined as:
\begin{equation}
\LL_{\rm S} = \left\langle 
\left| \rho^{\rm max}(i,j) - \rho^\sigma_{\P\widehat{\M}_b}(i,j) \right|
	\right\rangle_{i,j},
\label{eq:LS}
\end{equation}
where ${\rm Cov}\left(\cdot\right)$ and ${\rm Var}\left(\cdot\right)$ indicate the covariance and variance operators, respectively, $\left\langle\cdot\right\rangle$ denotes the average over the image, and being $\rho^{\rm max}$ a suitable upper bound CC map.
$\rho^{\rm max}$ may be just a constant 1, meaning that we seek to reach the maximum possible CC at each image location,
or a constant lesser than 1, relaxing a little the CC target, or be spatially varying, 
{\em e.g.} estimated at each location comparing the smoothed version of $\P$ with $\M_b$ (resized).
Details about the setting of $\beta$, $\sigma$, and $\rho^{\rm max}$, will be provided in the experimental section.

\section{Data, quality assessment, and methods}
\label{sec:data}

The main goal of the proposed work was to develop a new data-driven method for HS pansharpening that can outperform the state-of-the-art recently presented in the paper on the HS pansharpening challenge at IEEE WHISPERS 2022~\cite{Vivo23}. Leveraging on this claim, our experiments relied upon datasets, quality assessment procedures, and comparative methods exploited in the above-mentioned challenge and briefly described in the rest of this section. Besides, to enrich the comparative assessment, four additional deep learning solutions have also been enclosed \cite{He19, He20}.

\subsection{Datasets}
Despite the development of new approaches for HS pansharpening, most of them have been tested on simulated data neglecting an assessment using real data at full resolution. 
To overcome this limitation, PRISMA data have been distributed after the end of the WHISPERS challenge. Four datasets (the ones used for the contest) both at reduced and full resolutions have been shared. Each dataset comprises a PAN component and a HS image. The spatial resolution of the PAN image is 5 m. Instead, the HS sensor acquires about 250 spectral bands with a spatial resolution of 30 m. Before the announcement of the contest, only very few works on HS pansharpening of PRISMA images have been published. An application-oriented work using pansharpened PRISMA data has just been presented in 2021 \cite{Krem21}. Thus, the goal of the challenge has been to boost the research on HS pansharpening pushing researchers towards using new data, thus addressing new challenges. For example, the trade-off between computational cost (critical for images with hundreds of bands) and fusion performance or other peculiarities related to the HS pansharpening problem, 
such as the scale ratio different from 4 (that is instead widely used for multispectral pansharpening), 
the effects of a residual space-varying registration error between PAN and HS images, 
and the fusion of an elevate and sensor-dependent number of bands, which sometimes show low signal-to-noise ratios. 
Four teams accepted the challenge proposing innovative solutions relied upon machine learning and variational optimization-based methodologies. Despite of the use of state-of-the-art techniques, the four participating teams did not get outstanding results compared to the baseline and, for this reason, the organizing committee decided to close the contest and claim it is inconclusive (no winner).

In Tab.~\ref{tab:datasets}, some characteristics of the images are reported, while Fig.~\ref{fig:datasets} shows the data of the challenge. More specifically, four datasets are distributed, i.e., FR1 and FR2 for the assessment at full resolution, and RR1 and RR2 for the assessment at reduced resolution. In this work, a further dataset (i.e., FR0) has been exploited for validation purposes and to generate the initial weights of the proposed model.

\newcommand{\image}[1]{\includegraphics[width=0.23\linewidth]{./figures/prisma_img/#1.jpg}}
\begin{figure*}[t]
    \centering \scriptsize
    \setlength\tabcolsep{1pt}
    \begin{tabular}{cccc}
        \image{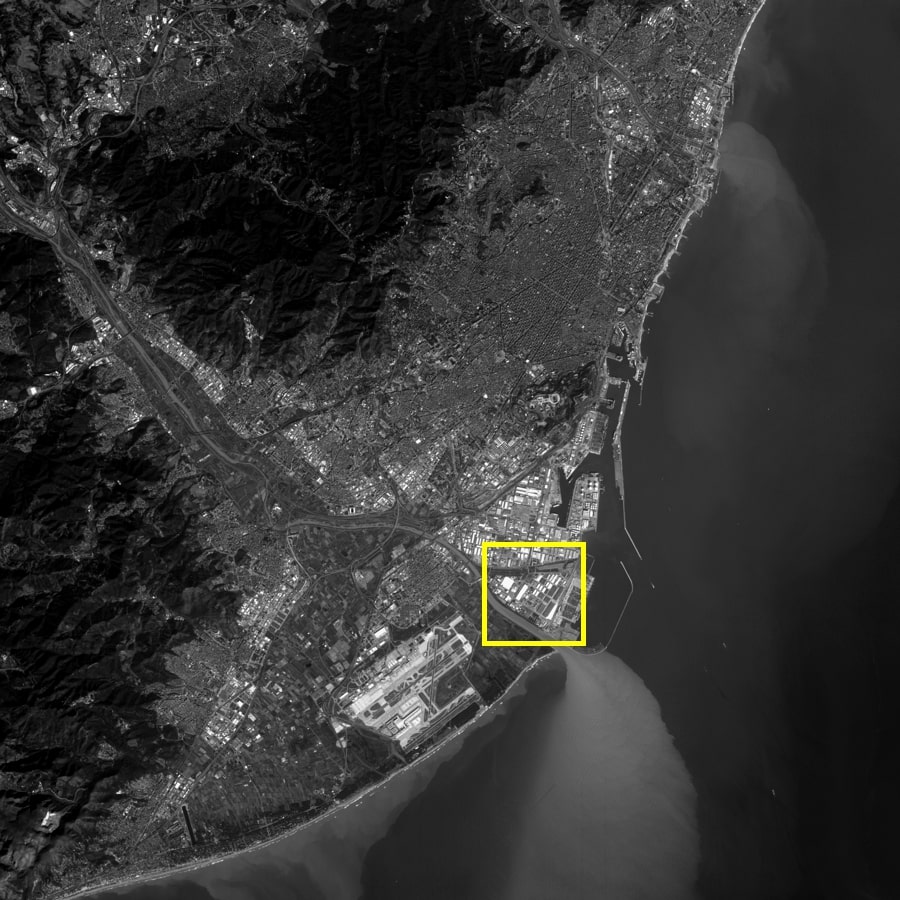}  & \image{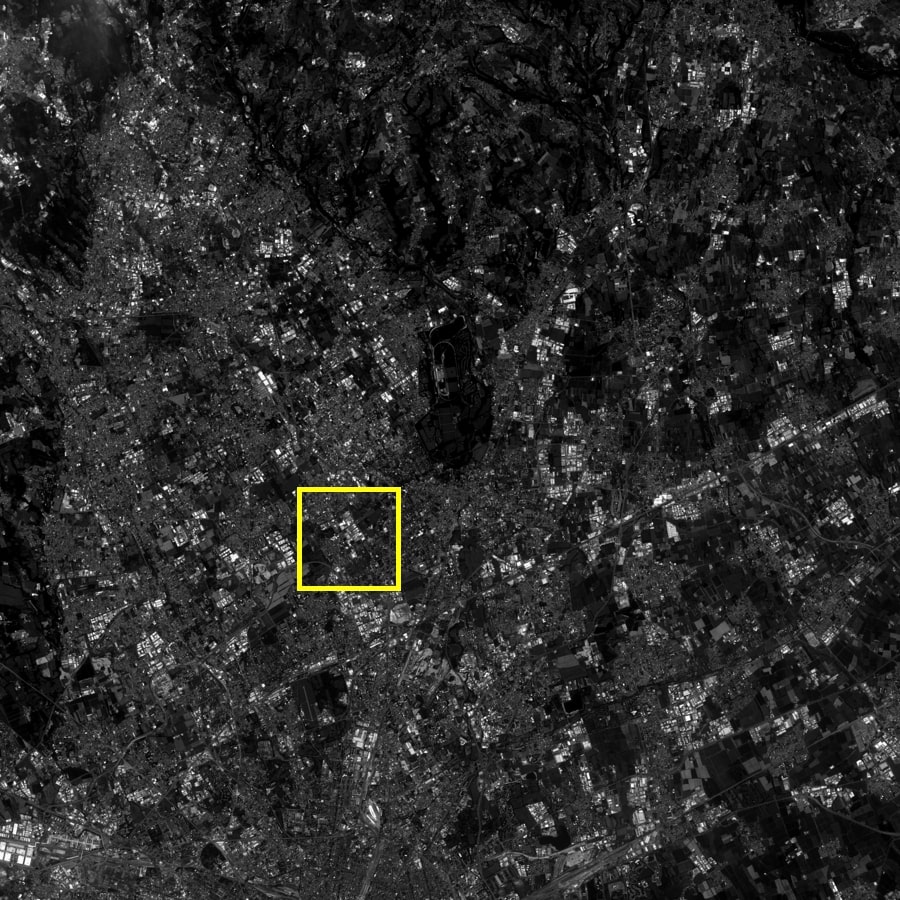} & \image{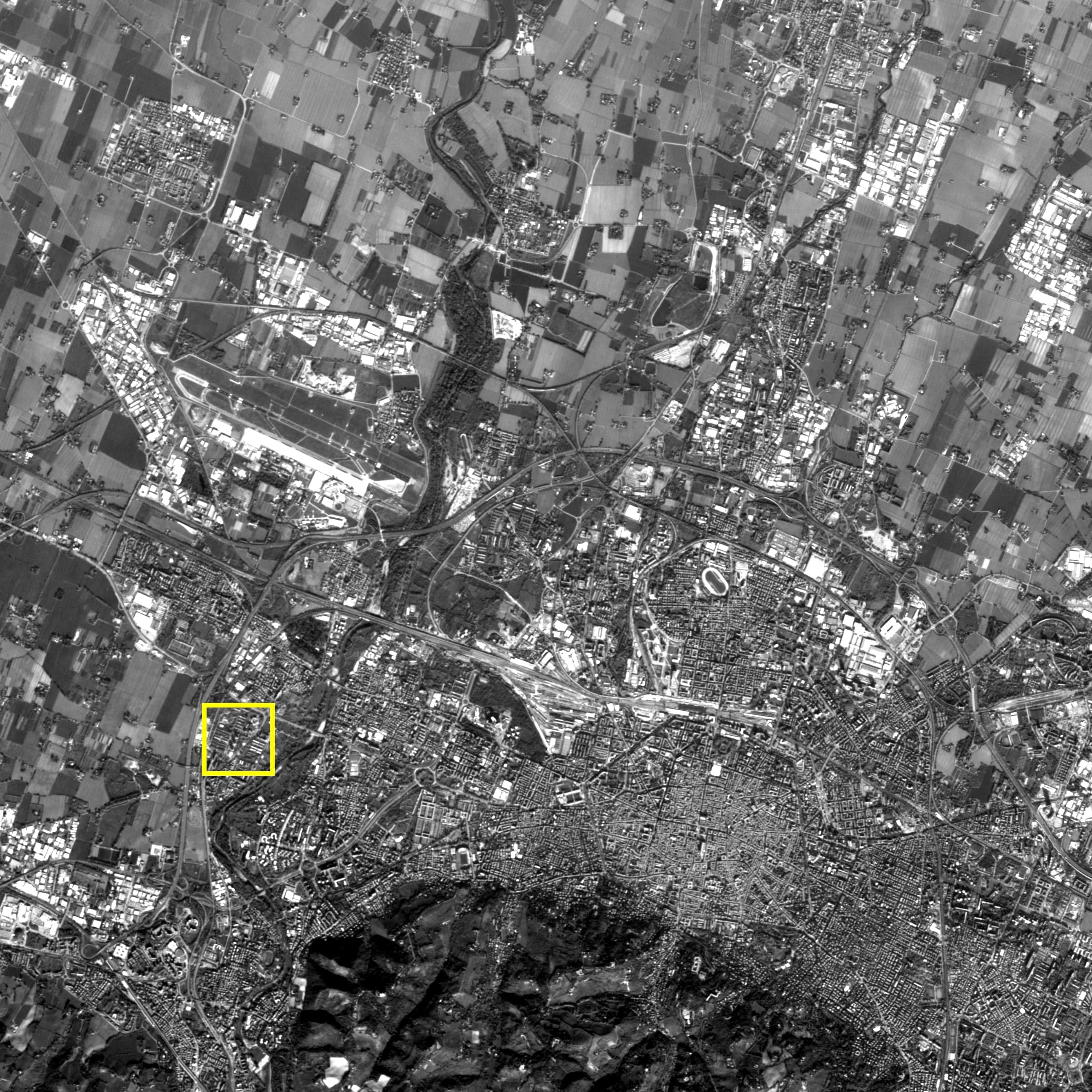} & \image{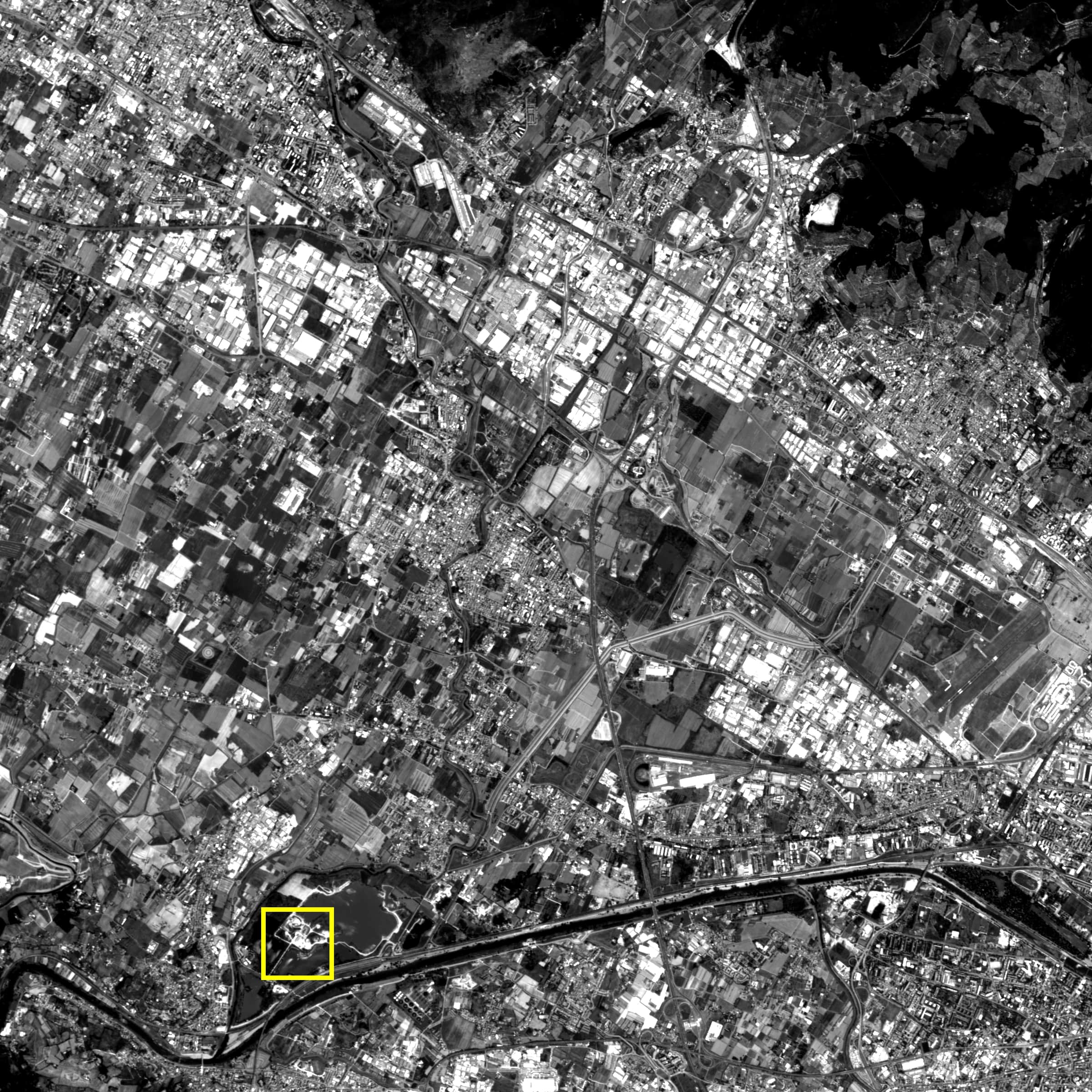} \\ 
        \multicolumn{4}{c}{PAN} \\[2mm]
        \image{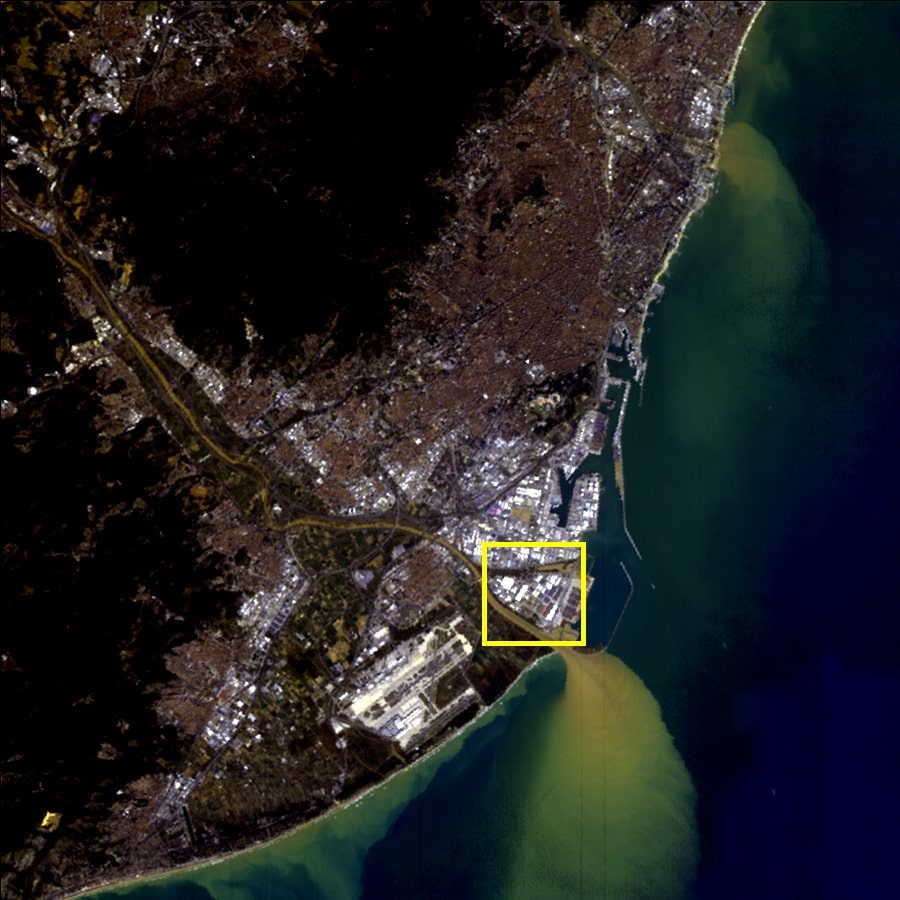}  & \image{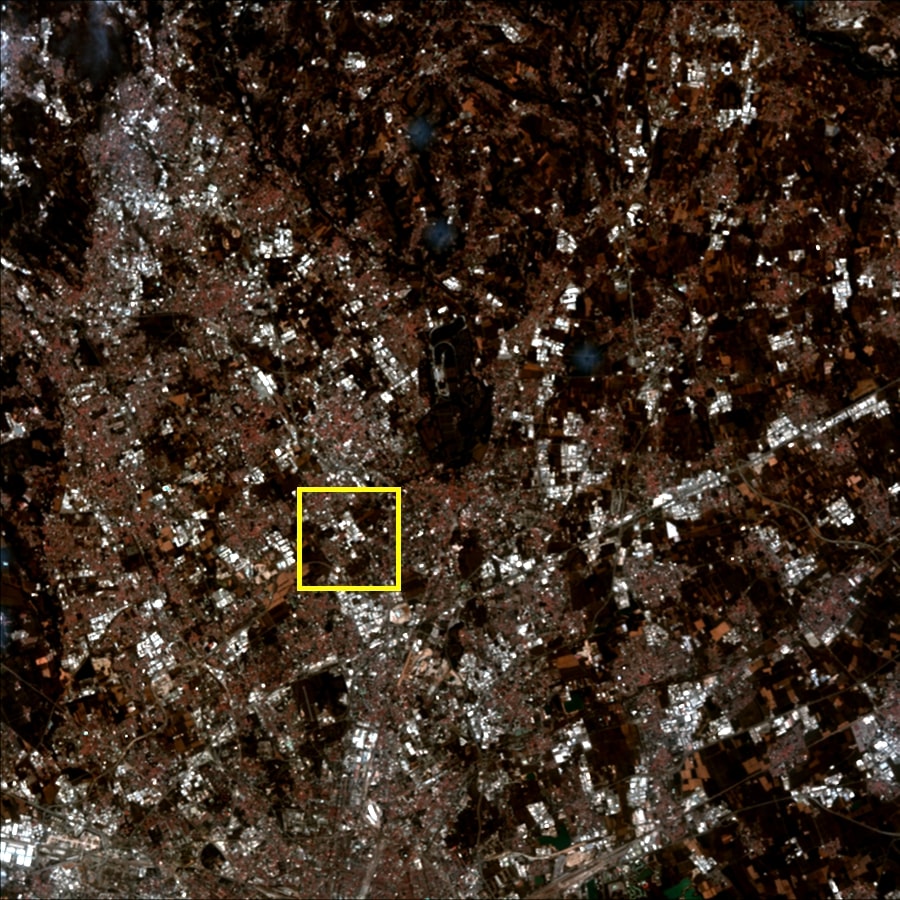} & \image{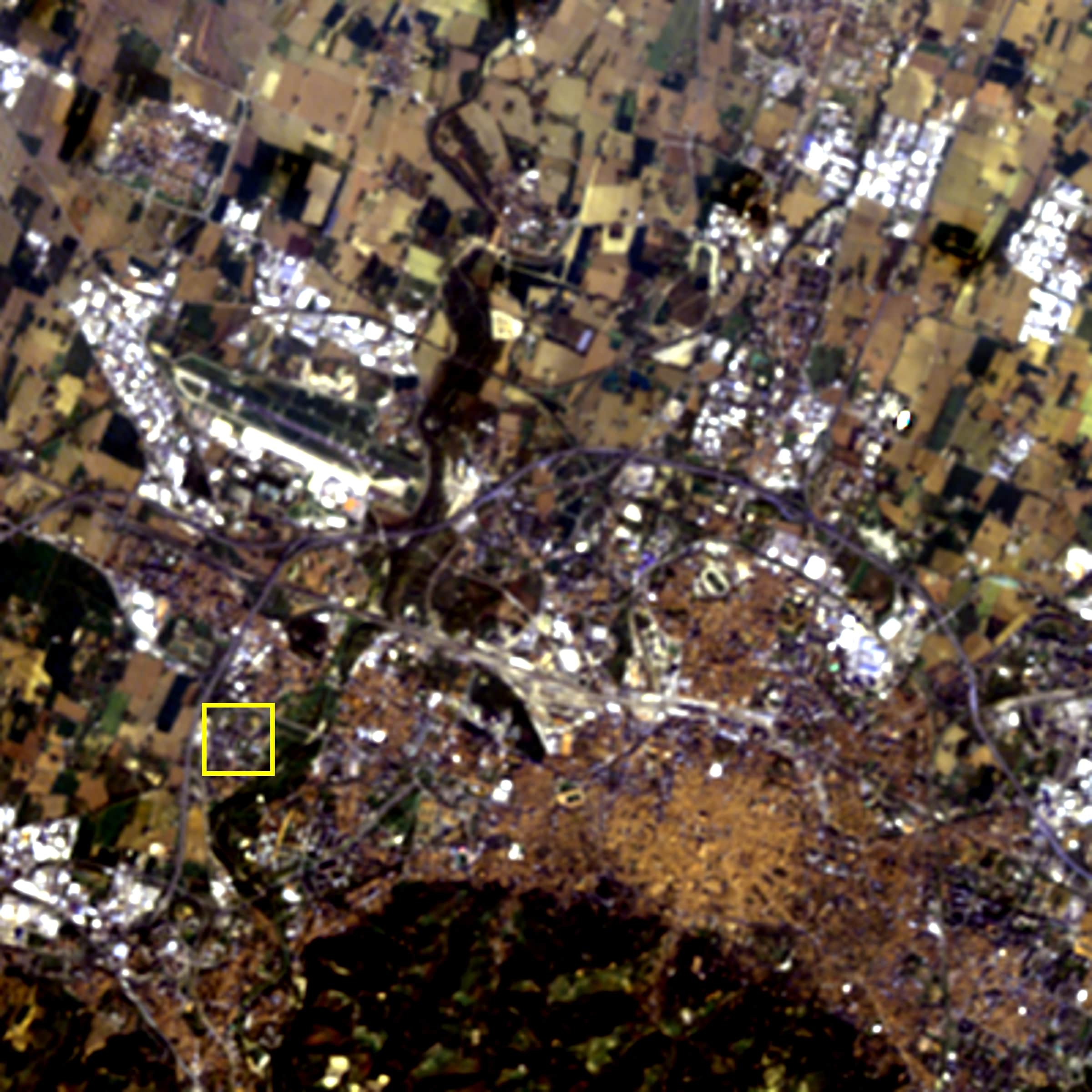} & \image{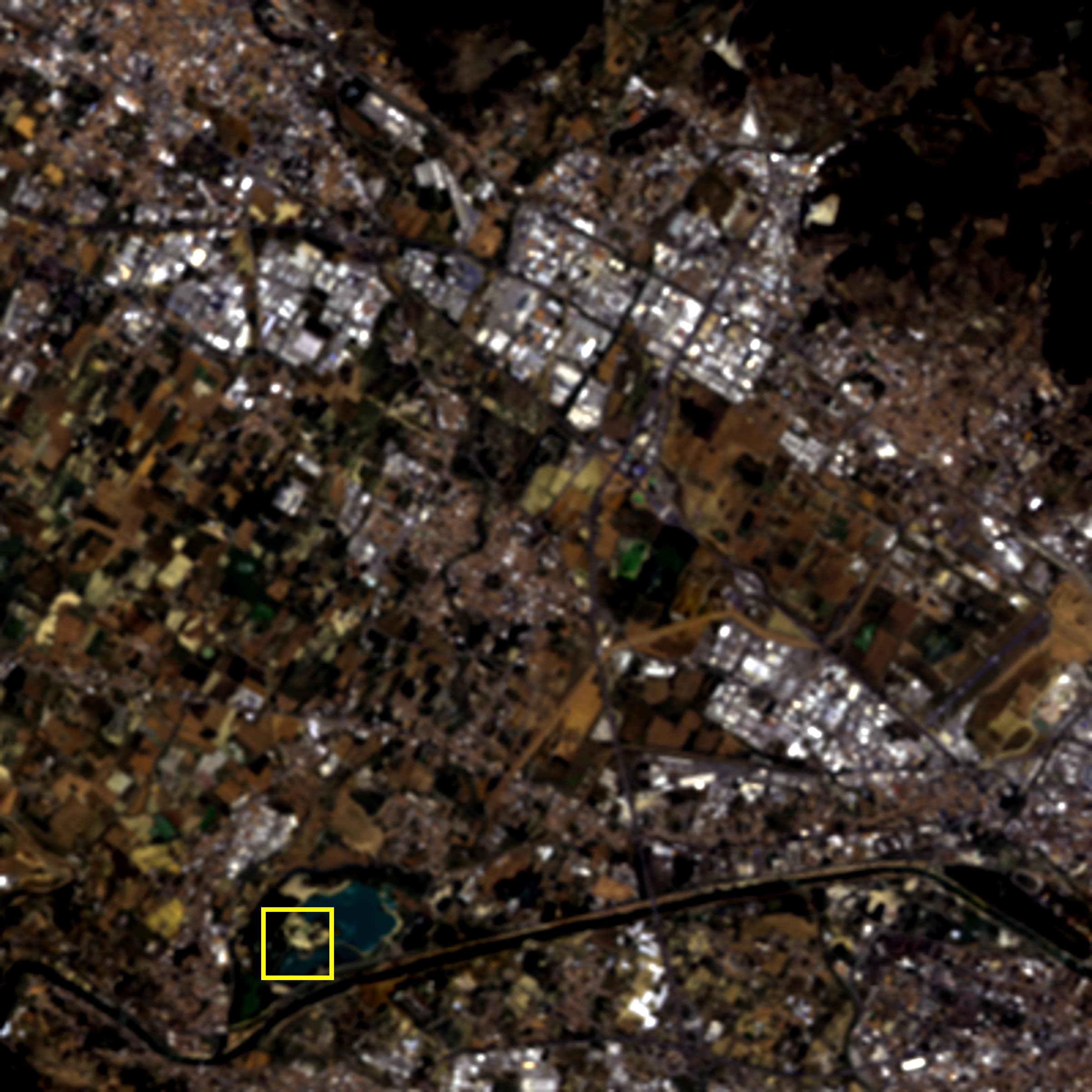} \\ 
        RGB wavelen. [nm] ($\lambda$): 660, 588, 442 &
		RGB $\lambda$: 632, 500, 434 & 
        RGB $\lambda$: 660, 588, 442 & 
        RGB $\lambda$: 679, 588, 456 \\[2mm]
        \image{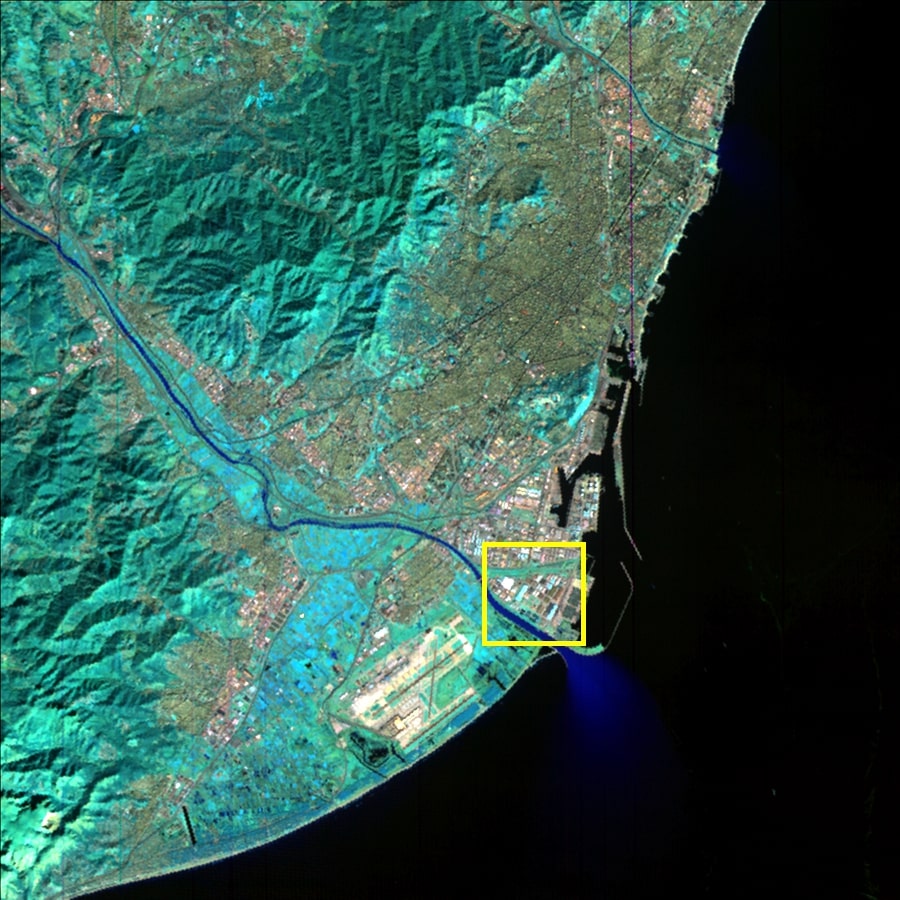}  & \image{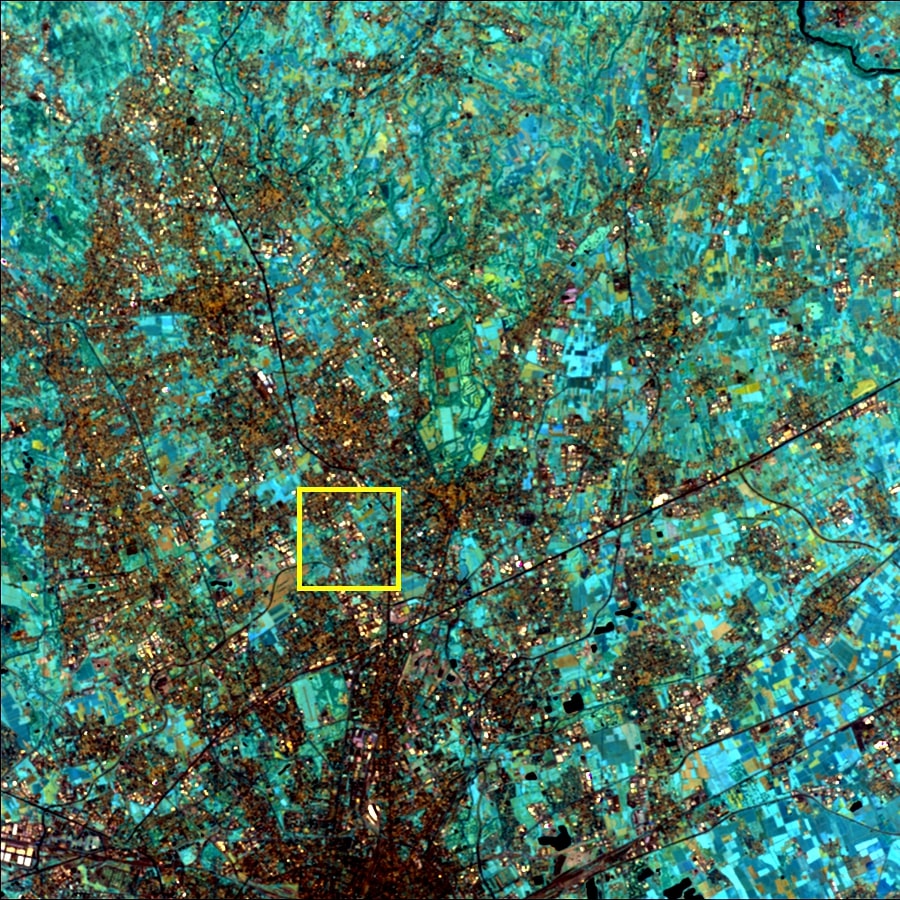} & \image{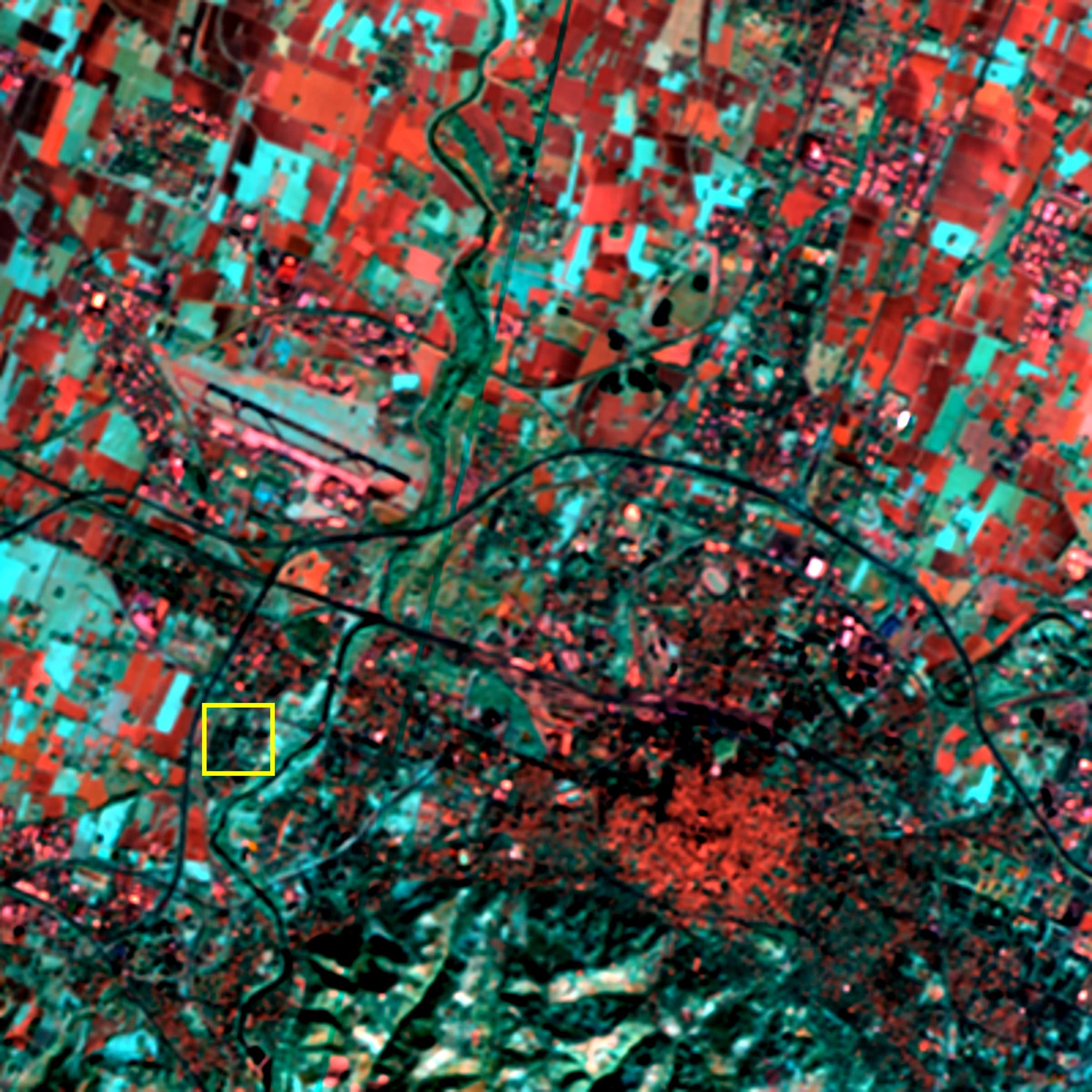} & \image{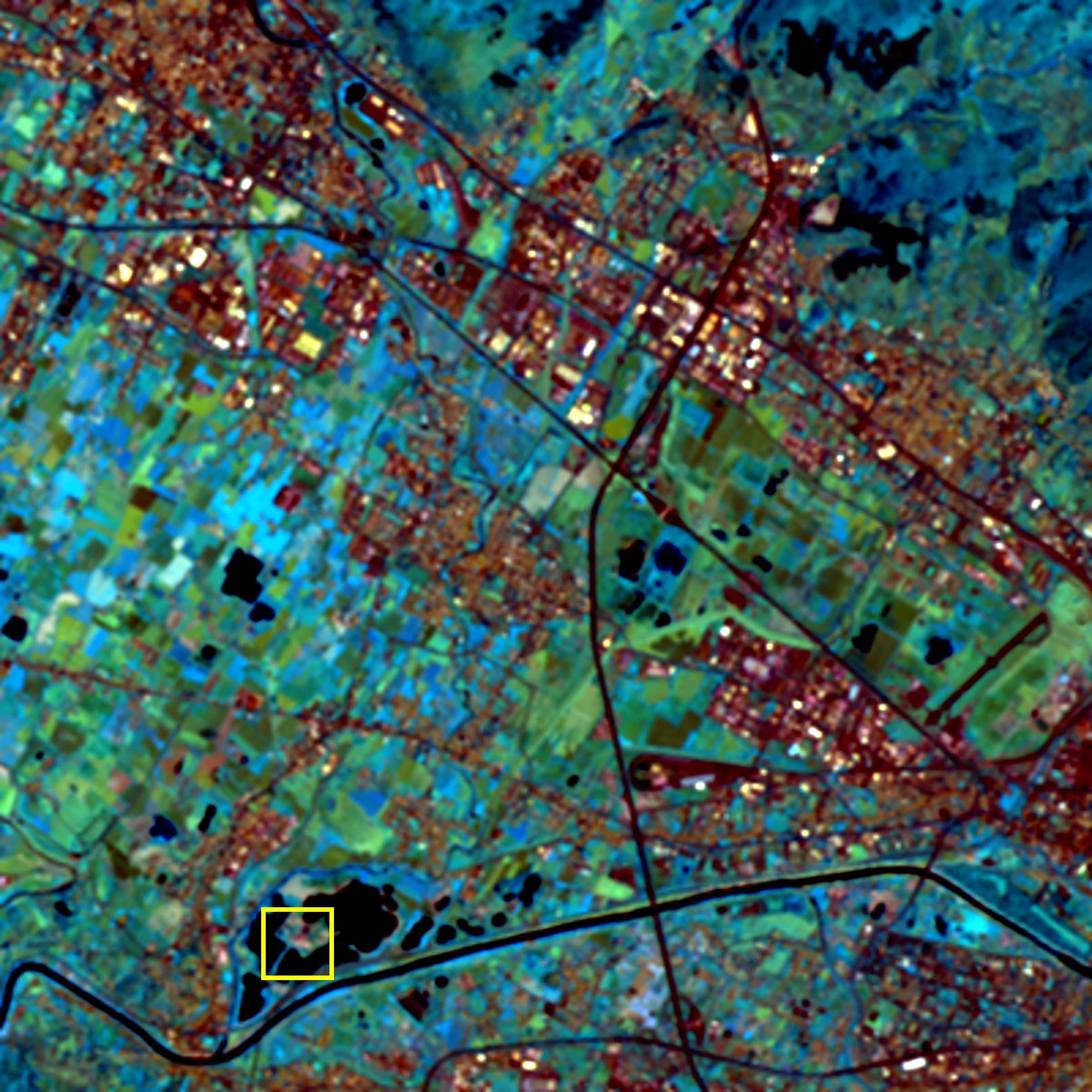} \\ 
        False-color $\lambda$: 2053, 1229, 770 &
        False-color $\lambda$: 1726, 1251, 750 & 
        False-color $\lambda$: 1586, 1038, 760 & 
        False-color $\lambda$: 2078, 1284, 802 
    \end{tabular}    
    \caption{The HS pansharpening challenge datasets \cite{Vivo23}.
    For each dataset (column), the PAN (top), a RGB subset using bands from the visible spectrum (middle), and a false-color subset using far apart bands (bottom) are shown. From left to right: RR1, RR2, FR1, and FR2.}
    \label{fig:datasets}
\end{figure*}

\begin{table}[t]
\centering
\begin{tabular}{ccccc}
\hline
Name & Zone & GSD [m] & PAN size & Bands \\
\hline
FR0 & Prato     & 5 & $2400\times2400$ & 73  \\
FR1 & Bologna   & 5 & $2400\times2400$ & 69  \\
FR2 & Florence  & 5 & $2400\times2400$ & 63  \\
RR1 & Barcelona & 30 & $900\times900$ & 59  \\
RR2 & Milan     & 30 & $900\times900$ & 73  \\
\hline
\end{tabular}
    \vspace{1mm}
\caption{Datasets. FR0 is used for validation purposes only.}
\label{tab:datasets}
\end{table}

\subsection{Accuracy indexes}
Assessing the performance of image fusion products is still an open issue, given the lack of full-resolution ground-truths. A widespread approach is to rely on the so-called synthesis property of Wald's protocol~\cite{Wald97}. The implementation of the above-mentioned protocol is based on a proper downsampling of the available data, under the hypothesis of invariance among scales of pansharpening algorithm performance~\cite{Vivo21b}. Hence, the original HS data play the role of ground-truth on which to measure the similarity with the fused product obtained by combining the degraded versions of the original PAN and HS images. The higher the similarity, the better the performance. This similarity degree can be evaluated through multidimensional score indexes~\cite{Vivo21b}:
\begin{itemize}
    \item The $Q2^{n}$~\cite{Garz09} is the multidimensional extension of the Universal Image Quality Index (UIQI)~\cite{Wang2002}. The upper bound of the index is one, even representing the optimal value.
		
    \item The Spectral Angle Mapper (SAM)~\cite{Yuha92} determines the spectral similarity (usually in degree) between the fused and the reference spectra. It is measured pixel-by-pixel and averaged over the whole image. The optimal value is zero. 
    \item ERGAS~\cite{Wald02} is a French acronym that stands for \textit{Erreur Relative Globale Adimensionelle de Synthèse} (dimensionless global relative error of synthesis). It is a normalized dissimilarity index (multidimensional extension of the root mean square error) that measures the radiometric distortion of the fused product with respect to the reference (ground-truth) image. The optimal value is zero.
	\item PSNR, measured in decibels, stands for Peak Signal-to-Noise Ratio and is one of the most popular quality index in the general image processing domain. Higher PSNR values indicating better quality.
\end{itemize}

It is worth pointing out that the reduced resolution assessment relies upon the invariance among scales hypothesis. This assumption could not be valid. Furthermore, the accuracy can depend on how to degrade the original PAN and HS products~\cite{Vivo21b}. As a consequence, to provide a complete assessment of pansharpening algorithms, the validation at full resolution is also adopted~\cite{Shah2008, Zhou1998, Alpa08, Khan09, Vivo18, Scar2022}. In this paper, we followed the indications in~\cite{Vivo23} using the same quality indexes at full resolution. More specifically, the $Q^{*}$ index~\cite{Arie22} is exploited consisting of a spectral distortion index~\cite{Khan09, Arie22}, $D_{\lambda}$, based on Wald's consistency property~\cite{Wald97}, and a regression-based spatial distortion index, $D_{S}$, firstly proposed in~\cite{Alpa18} and then deeply investigated in~\cite{Arie22}. In the ideal case, both the spatial and spectral distortion indexes are zero, thus obtaining a $Q^{*}$ index equal to one.

For additional details about the PSNR the interested readers can refer to \cite{Yim2011} while, 
for all other reduced and full resolution indexes, implementation details are given in~\cite{Vivo23}
and in the related freely available toolbox\footnote{Website: \url{https://openremotesensing.net/}}.

\begin{table}[t]
\centering
\setlength{\tabcolsep}{2pt}
\begin{tabular}{p{7.8cm}}
\hline
\multicolumn{1}{c}{\rule{0pt}{7pt}\bf Component Substitution (challenge baseline)} \\
GS \cite{Labe00}, GSA \cite{Aiaz07} \\ \hline
\multicolumn{1}{c}{\rule{0pt}{7pt}\bf Multiresolution Analysis (challenge baseline)} \\
AWLP \cite{Otaz05}, MTF-GLP \cite{Aiaz06,Alpa17}, MF \cite{Rest16} \\ \hline
\multicolumn{1}{c}{\rule{0pt}{7pt}\bf Challenge competitors: Variational or Machine/Deep Learning} \\
Team 1 \cite{Vivo23}, Team 2 \cite{Vivo23}, Team 3 \cite{Vivo23}, Team 4 \cite{Vivo23}\\ \hline
\multicolumn{1}{c}{\rule{0pt}{7pt}\bf Deep Learning}\\
HyperPNN1 \cite{He19}, HyperPNN2 \cite{He19}, HSpeNet1 \cite{He20}, HSpeNet2 \cite{He20}\\ \hline
\end{tabular}
    \vspace{2mm}
	\caption{Benchmarking approaches.}
\label{tab:methods}
\end{table}

\subsection{Benchmarking}
Tab.~\ref{tab:methods} summarizes the techniques used in our experimental analysis.
The benchmarking approaches taken from the WHISPERS challenge are described in~\cite{Vivo23}. More specifically, five methods, exploited as baseline solutions for the challenge, are borrowed from the pansharpening literature.  The first two~\cite{Vivo21, Vivo15a} belong to the Component Substitution (CS) class, i.e., the Gram-Schmidt (GS)~\cite{Labe00} approach and its adaptive version, GSA~\cite{Aiaz07}. The other three baseline methods are representative of the Multi-Resolution Analysis (MRA). More in detail, the third method is the classical Additive Wavelet Luminance Proportional (AWLP)~\cite{Otaz05}; the fourth technique is the MTF-Generalized Laplacian Pyramid (MTF-GLP)~\cite{Aiaz06} with histogram-matching~\cite{Alpa17}; finally, the last baseline method is the Morphological Filters (MF)~\cite{Rest16}. The other four techniques (labeled as Team 1 to 4) are innovative variational optimization-based and machine learning-based solutions proposed by the participants to the hyperspectral pansharpening challenge~\cite{Vivo23}.
Finally, four extra-challenge deep learning solutions \cite{He19, He20} have been reimplemented and trained on PRISMA data for further comparison.
In particular, for these methods, lacking extra datasets for training and giving the heterogeneity (different numbers of bands) of the test datasets,
we used in training a portion of the same test datasets, applying the canonical resolution downgrade protocol needed for supervised models. 
\section{Experimental validation}
\label{sec:validation}
In this section we show and discuss several experimental results aimed 
to support our design choices.
In particular, 
we will analyze the relationship between the proposed loss and the accuracy indicators, \ref{sec:loss}.
We will show the impact of the model propagation mechanism, \ref{sec:propagation}, and of the tuning, \ref{sec:tune}.
Then, we will provide details about the setting of the spatial loss term, \ref{sec:spatial}.
Finally, we carry out an ablation study on the network architecture, \ref{sec:ablation},
before concluding with some details about the pretraining phase, \ref{sec:pretrain}.

\subsection{Loss and accuracy}
\label{sec:loss}
\begin{figure}
\centering
\includegraphics[scale=1.2]{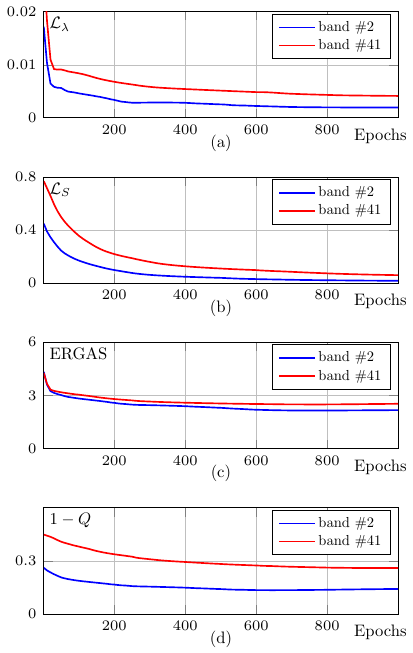}
\caption{Training curves for single-band pansharpening on RR1. 
Spectral (a) and spatial (b) loss terms, ERGAS (c), and $1-Q$ (d).}
\label{fig:consistency_RR}
\end{figure}

The first experiment deals with the choice of the loss. 
Since we propose an unsupervised one, 
it is not obvious or, at least, not always possible, 
that the smaller the loss the better the accuracy, according to the available quality indicators. 
Therefore, it is a fundamental question to understand to what extent there is 
agreement between the loss and the quality indicators.
Toward this goal we have selected a full-resolution (FR2) and
a reduced-resolution (RR1) dataset, 
restricting the pansharpening to a single HS band.
For both datasets we consider two opposite conditions: a highly and a weakly correlated (to the PAN) HS band.
Since we target a single HS band, 
the proposed solution reduces to the traditional pansharpening without the need of model propagation.
Moreover, to avoid eventual biases, 
we run the fine-tuning target adaptation from scratch (random initial weights).
In particular, 
here we are interested to monitor the evolution of the loss in comparison with the evolution of the 
accuracy indicators during the training.
In Fig.~\ref{fig:consistency_RR}, all the involved training curves are gathered for the reduced-resolution case:
spectral (a) and spatial (b) loss terms, ERGAS (c), and $1-Q$ (d).
SAM is not applicable in the single-band case, hence it is not monitored.
Blue and red curves refer to the cases of highly (\#2) and weakly (\#41) correlated bands, respectively.
As first observation, 
we notice that both the loss terms, $\LL_\lambda$ and $\LL_S$, keep descending monotonically 
even if they seem to be very close to their lower bound after about thousand iterations.
To this regard, it should be remarked that the spatial and the spectral loss terms are partially fighting with each other \cite{Ciot2022}
and could be very hard to bring both at zero. 
Such a tradeoff is more noticeable for band \#41 where a smaller correlation with the PAN
makes it difficult to minimize $\LL_S$ without sacrificing spectral consistency ($\LL_\lambda$).
Moving the focus on ERGAS,
we can observe a coherent behavior with both the loss terms, 
as it decreases monotonically at least in the first 600 iterations,
before reaching a plateau. 
This occurs for both the bands and a similar behavior is registered for $1-Q$.
Eventually, it seems safe to say that the two chosen loss terms are generally consistent 
with the standard quality indicators ERGAS and $Q$.
Above certain quality levels (after 600 iterations), 
they seem to loose their correlation with ERGAS and $Q$. 
But this is somewhat expected as they are no-reference indicators
whereas the latter are reference-based.

\begin{figure}
\centering
\includegraphics[scale=1.2]{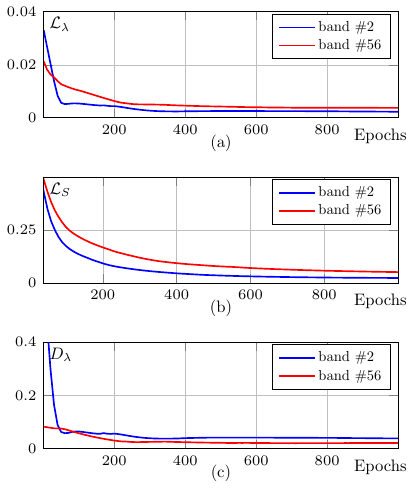}
\caption{Training curves for single-band pansharpening on FR2. 
Spectral (a) and spatial (b) loss terms, and $D_\lambda$ (c).}
\label{fig:consistency_FR}
\end{figure}

Let us now move the focus on the most interesting full-resolution case 
with the help of Fig.~\ref{fig:consistency_FR} that shows the related training curves:
spectral (a) and spatial (b) loss terms, and the spectral distortion index $D_\lambda$ (c). 
We do not show the spatial distortion $D_S$ as it loses its meaning if computed on a single spectral band.
Again, blue and red curves refer to the cases of highly (\#2) and weakly (\#56) correlated bands, respectively.
In this case the analysis is simpler as we can relate homogeneous no-reference quantities.
Focusing on the spectral numerical figures, 
we can observe a perfect agreement between $D_\lambda$ and $\LL_\lambda$,
no matter which band is concerned.
On the other side,
we do not dispose of a numerical reference to judge the spatial consistency of $\LL_S$.
Nonetheless,
we can first observe that $\LL_\lambda$ and $\LL_S$ can be minimized simultaneously
and, at the same time, they show different trajectories, 
symptomatic of a weak correlation between the two,
with $D_\lambda$ clearly linked to $\LL_\lambda$ rather than to $\LL_S$.
Moreover,
with the help of Fig.~\ref{fig:focusing}, 
we can appreciate (subjectively)
the improvement of the spatial quality due to the minimization of $\LL_S$ 
through the visual inspection of some sample results obtained 
along the tuning process (full-resolution case).
\newcommand{\ima}[1]{\includegraphics[width=0.28\columnwidth]{./figures/progress/FR2_#1.png}}
\begin{figure*}
\setlength{\tabcolsep}{1.5pt}
\centering
{\footnotesize
\begin{tabular}{ccccccc}
\ima{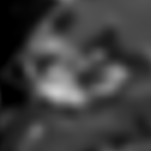} & \ima{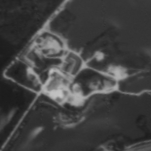} & \ima{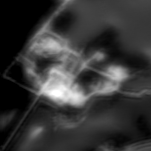} & \ima{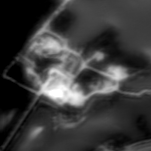} & \ima{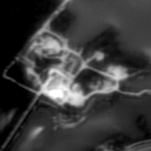} & \ima{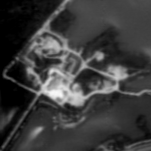} & \ima{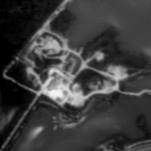} \\[1mm]
LR band \#2 & PAN & \#it. = 50 & \#it. = 100 & \#it. = 200 & \#it. = 400 & \#it. = 1000  \\
& & 						$\LL_\lambda=0.0057$ & $\LL_\lambda=0.0054$ & $\LL_\lambda=0.0044$ & $\LL_\lambda=0.0025$ & $\LL_\lambda=0.0023$  \\
& & 						$\LL_S=0.349$ & $\LL_S=0.175$ & $\LL_S=0.093$ & $\LL_S=0.047$ & $\LL_S=0.025$ \\[4mm]
\ima{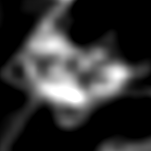} & \ima{PAN} & \ima{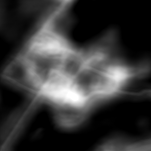} & \ima{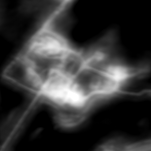} & \ima{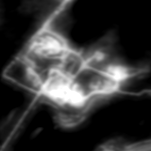} & \ima{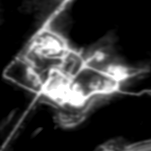} & \ima{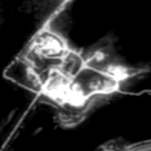} \\[1mm]
LR band \#56 & PAN & \#it. = 50 & \#it. = 100 & \#it. = 200 & \#it. = 400 & \#it. = 1000  \\
 & & 						$\LL_\lambda=0.0127$ & $\LL_\lambda=0.0103$ & $\LL_\lambda=0.0063$ & $\LL_\lambda=0.0046$ & $\LL_\lambda=0.0037$ \\
& & 						$\LL_S=0.430$ & $\LL_S=0.264$ & $\LL_S=0.169$ & $\LL_S=0.094$ & $\LL_S=0.052$ 
\end{tabular}
}
\caption{Single-band sample pansharpening results during training.}
\label{fig:focusing}
\end{figure*}
From left to right are shown the original low-resolution HS bands, the PAN, and a series of pansharpening results progressively singled out along the training process.
Below each result is reported the number of tuning iterations, and the values of the corresponding spectral and spatial loss terms.
The spectral adherence of the these outcomes to the interested low-resolution band is hard to see. 
However, the spectral distortion $D_\lambda$ decreases monotonically (Fig.~\ref{fig:consistency_FR}~(c)),
guaranteeing for a progressive spectral quality enhancement.
Indeed, it is worth noticing that the most of the spectral distortion (see $\LL_\lambda$ or $D_\lambda$) 
is removed in the first 50 iterations for band \# 2 (first 200 iterations for band \#56),
whereas the spatial loss presents a more distributed and regular decay.
This allows us to easily attribute to $\LL_S$ the spatial improvement 
clearly visible in the sequence of Fig.~\ref{fig:focusing}.
To be more specific,
we can look at the pansharpening results for band \#2 at 50 and 100 iterations.
In this interval, the spectral loss remains nearly constant (0.0057 to 0.0054)
whereas the spatial one halves (0.349 to 0.175).
Besides, the visual inspection of these partial results reveals a remarkable improvement,
which cannot but depend on the reduction of $\LL_S$.
Similar observations can be done for band \#56 moving from 400 to 1000 iterations.
Encouraged by this experimental analysis, 
we decided to keep the proposed spatial loss (Eq.\ref{eq:LS}), that ensures a consistent behavior according to a visual assessment.

\subsection{Spectral correlation analysis and model propagation}
\label{sec:propagation}

\begin{figure}
\centering
\includegraphics[scale=1]{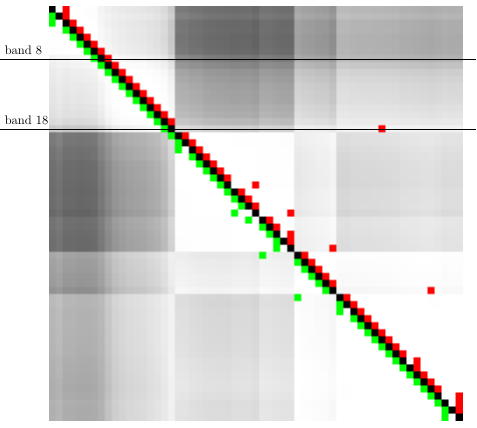}
\caption{Covariance matrix for a sample HS PRISMA datacube. 
For each band (fix a row index) the most correlated previous and next bands are marked in green and red, respectively.}
\label{fig:cc}
\end{figure}
To gain insight the inter-band dependence, 
let us have a look at the covariance matrix normalized in the range [-1,1] (correlation coefficient) shown in Fig.~\ref{fig:cc}
associated to a sample HS PRISMA image.
The maximum values are on the diagonal and marked in black. 
For each given band ({\em e.g.} look at row \#8), the best correlated bands looking backward and forward are marked in green and red, respectively.
Focusing on the backward case, we can notice that, 
in the large majority of the cases, 
the most correlated band is just the previous one.
When this is not the case, the correlation value for that band is, however, very close to the maximum. 
In the forward search, we have nearly the same situation with just one exception for band 18.
We have carried out the same analysis on all the available datasets and the conclusions were always the same.
Based on the above considerations, it makes perfectly sense to propagate the model from one band to the next one (or previous one, if we proceed in the opposite direction).

\begin{figure}
\centering 
\includegraphics[scale=1.2]{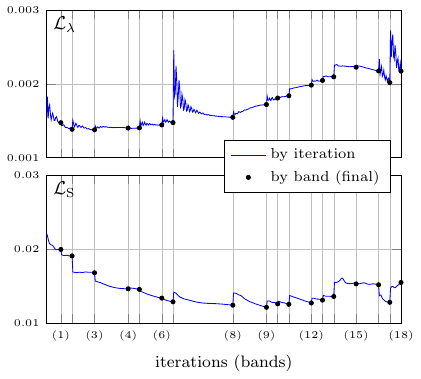}
\caption{Concatenated spectral (top) and spatial (bottom) loss progress during model propagation (close-up on the first 18 bands).
Each vertical stripe corresponds to a different band whose final loss is marked with a dot.
}
\label{fig:loss}
\end{figure}
To further clarify the propagation tuning process, 
in Fig.~\ref{fig:loss}, it is shown the progress of the loss adaptation, limited to a subset (1-18) of bands for ease of view.
The loss (Sec.~\ref{sec:unsup loss}) comprises two terms responsible for the spectral (top) and spatial (bottom) consistencies.
Checkpoints highlight the final loss for each band.
A careful inspection of the curves reveals that in many cases, 
the same model for a given band already fits well (for $\LL_\lambda$, $\LL_S$, or both) on the next band when a new tuning starts.
In some case, for example on band 8, we can notice a typical behavior, 
where the propagated model needs to be adjusted for the new band.
In fact,
coming back to Fig.~\ref{fig:cc},
we can observe that the correlation between adjacent bands keeps very high until band 7.
Then, a drop is registered for band 8 because of a larger spectral gap,
which justifies the model mismatch between bands 7 and 8.

It is also worth to observe how spectral and spatial losses contrast each other when both reach too small values
(how much small depends on the band and, in particular, on its correlation with the PAN).
In fact, in a few cases, for example for bands 9, 12, and 18, only one of the two losses decreases.
The spatial loss descends at a price of a little increase of the spectral loss for bands 9 and 12 
(notice that the spectral loss is one order of magnitude smaller than the spatial loss). 
For band 18, instead, likely because of a large mismatch with the preceding band,
the spectral loss dominates the tuning process. 

Finally, as general remark,
we observe that the balance between $\LL_\lambda$ and $\LL_S$ is function of the band
and of its (stronger or weaker) relationship with the PAN.

\begin{figure}
\centering
\includegraphics[scale=0.57]{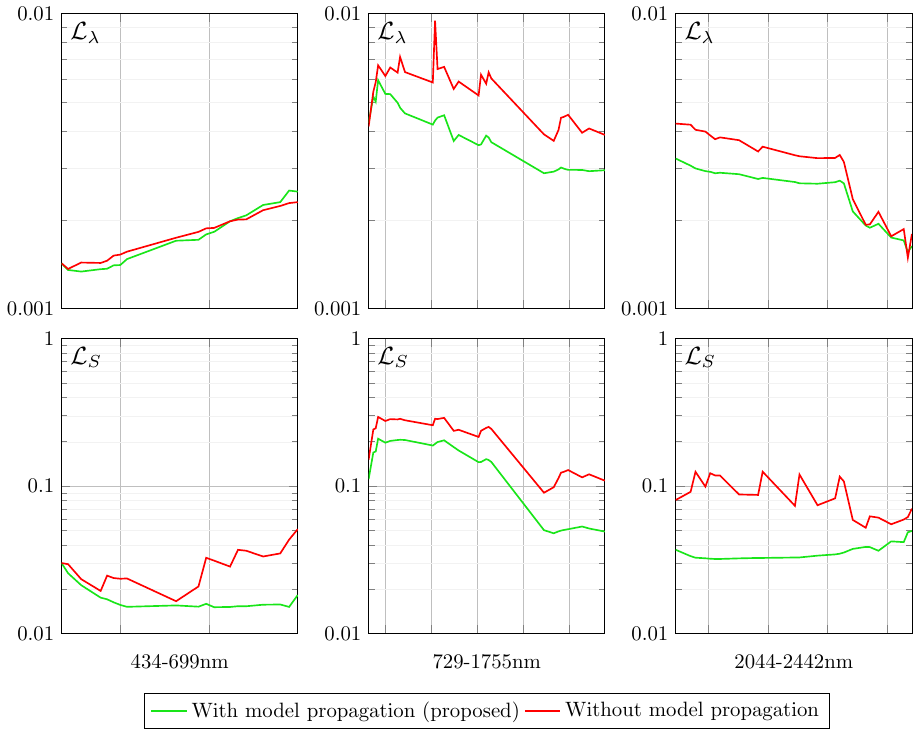}
\caption{Band-wise spectral (top) and spatial (bottom) loss components after tuning with (blue) or without (red) model propagation.}
\label{fig:propagation}
\end{figure}

Let us now focus on a set of validation experiments dealing with the effectiveness of the model propagation scheme.
Here, all spectral bands are concerned according with the processing chain summarized in Fig.~\ref{fig:rolling}.
First, we compare the proposed solution based on model propagation, 
where the $b$-th tuning block inherits as starting parameters
the ones ($\phi_{b-1}$) tuned on band $b-1$,
with the case where the model is not propagated and the tuning always starts from the same set of parameters, $\phi_0$.
In both the cases, the number of tuning iterations is computed band-wise using Eq.~\ref{eq:Nb}.
The experiment is run on the full-resolution image FR0 and the results are depicted in Fig.~\ref{fig:propagation} in terms of achieved band-wise loss
at the end of the tuning, 
separately for the spectral (top) and spatial (bottom) components.
Due to a non-uniform coverage of the spectral range by the available HS datasets,
the wavelength axis has been split in three intervals for a better visualization.
The plots show a clear gain due to the propagation of the model on both spectral and spatial sides 
and all along the wavelength axis, with quite large gaps especially in terms of $\LL_S$.

To further validate the proposed solution, we have also compared the proposed forward model propagation with the backward option
on the validation dataset FR0.
The resulting accuracy indicators are reported in Tab.~\ref{tab:backward}.
\begin{table}
	\centering
    \begin{tabular}{lccc}
        \hline
         & $D_\lambda$ & $D_S$ & $Q^\ast$ \\
         \hline
        Forward & 0.0294 &  0.0184 & 0.9527  \\
        Backward & 0.0319 & 0.0198 & 0.9489  \\
        \hline
    \end{tabular}
\label{tab:backward}
\vspace{1mm}
	\caption{Numerical comparison between the forward and backward propagation schemes on the FR0 dataset.}
\end{table}
The numbers show that the forward propagation option provides only slightly better scores. 
Experiments carried out on other datasets, however, confirm a substantial equivalence between the two solutions,
hence we eventually opted for the causal ordering (forward) without loss of generality.

\subsection{Tuning strength}
\label{sec:tune}
\begin{figure}
\centering
\includegraphics[scale=0.57]{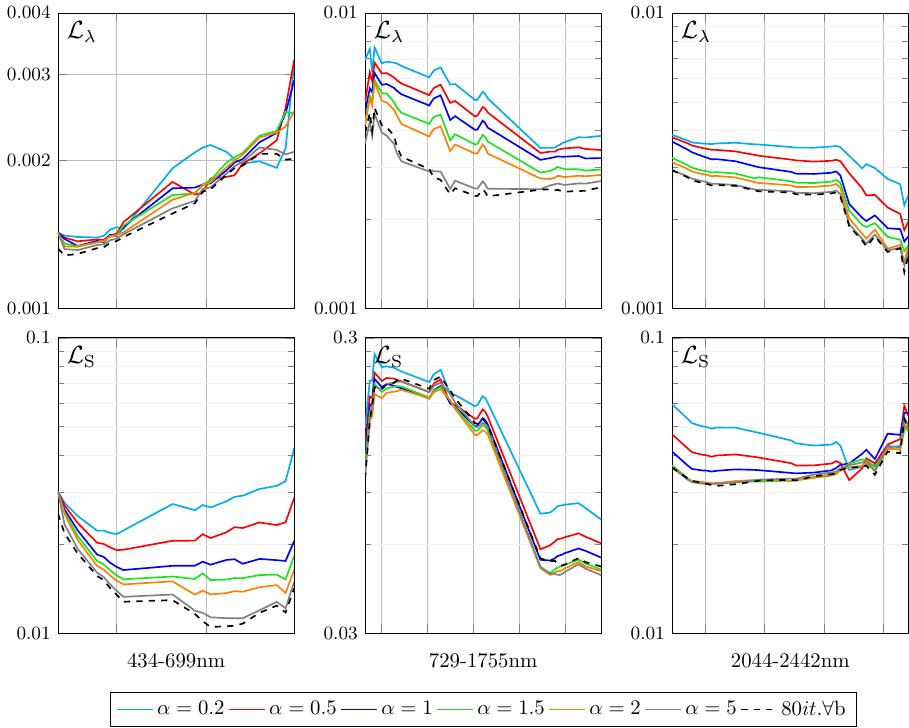}
\caption{Results obtained on the FR0 image. The final value of spectral (top) and spatial (bottom) losses against band wavelength for the different configurations of $\alpha$.}
\label{fig:alpha}
\end{figure}

According to the proposed empirical rule to fix the number $N_b$ of tuning iterations per band (Eq.~\ref{eq:Nb}),
such number is proportional (up to a maximum saturation value) to the wavelength distance from the previous band, $\Delta\lambda_b = \lambda_b-\lambda_{b-1}$.
In practice, since in most of the cases such a step is about 10nm (the minimum for PRISMA), 
a proportionality factor $\alpha = 1$~iteration/nm
would amount to run about 10 iterations per band except those (fewer) cases where larger spectral gaps are concerned.
Of course, the larger the $\alpha$, the better the fitting to the target image, but also, the higher the computational time.
Therefore, to fix an appropriate value for $\alpha$ we have run a sequence of validation tests on the FR0 image
using different values of $\alpha$ (0.2, 0.5, 1, 1.5, 2, 5). 
In addition, 
we also show the limit case where we perform the maximum number of iterations (80) for all bands, regardless of $\Delta_\lambda$.
On the one hand, we provide the band-wise value of the loss in Fig.~\ref{fig:alpha}, splitting, as usual, the spectral (top) and spatial (bottom) terms.
On the other hand, we give the computation time for each configuration in Tab.~\ref{tab:time}.

\begin{table}[htp]
    \centering
    \begin{tabular}{lc}
    \hline
     Setting & Computional time \\
    \hline
      $\alpha = 0.2$  & 10' \\
      $\alpha = 0.5$  & 22'\\
      $\alpha = 1.0$  & 41' \\
      $\alpha = 1.5$  & 51' \\
      $\alpha = 2.0$  & 1h02' \\
      $\alpha = 5.0$  & 1h33' \\
      $80 \mbox{it.} \forall b$  & 1h39' \\
    \hline
    \end{tabular}
        \vspace{1mm}
    \caption{Computation time to perform target adaptation on a $2400\times2400$ PRISMA image with 73 bands using a NVIDIA GeForce RTX 2080 Ti GPU with 11GB of memory.}
    \label{tab:time}
\end{table}
From Fig.~\ref{fig:alpha}, it can be observed that, 
with respect to the limit case (dashed) where the maximum number of iterations is run for all bands,
both the spectral (outside the visible spectral range only) and the spatial loss (in the visible range only)
register a progressive deterioration as $\alpha$ decreases. 
Besides, as reported in Tab.~\ref{tab:time},
smaller $\alpha$s provide quicker inference.
On the basis of these observations,
we decided to fix a tradeoff value $\alpha=1.5$ for testing the proposed solution,
leaving to the end user the possibility to change this default setting upon specific needs and hardware
(in our experiments we exploited an NVIDIA GeForce RTX 2080 Ti GPU with 11GB of memory).

\subsection{Spatial loss configuration}
\label{sec:spatial}
The proposed loss (\ref{eq:L}) comprises two contributes, 
the spectral (\ref{eq:LL}) and the spatial (\ref{eq:LS}) consistency terms.
While the former leverages on a standard regression error function, such as the $\ell_1$-norm,
and relates pretty well with spectral distortion, $D_\lambda$,
the latter is more complex and its contribute to the spatial quality that is less obvious.
To gain insight in its role, 
here we provide additional details on its configuration.
With respect to the classical pansharpening problem, 
the most critical difference is that
the PAN image spectrally overlaps only with a subset of HS bands.
Therefore, 
indiscriminately forcing correlation between each pansharpened band and the PAN may be detrimental.
To mitigate eventual side effects,
on one hand, we halve (from 0.5 to 0.25) the weighting hyper-parameter $\beta$ (\ref{eq:L}) 
for those HS bands which do not overlap with the PAN,
on other hand, we use a spatial-spectral varying correlation bound, $\rho^{\rm max}$ (\ref{eq:LS}).
To estimate this bound we resort to a downgrading process (just a low-pass filtering) applied to the PAN.
The HS image is upsampled to the PAN scale
and then the bound local correlation coefficient is computed.
However, although the involved images have the full target size,
the lack of high-frequency content would make meaningless the computation of the correlation coefficient 
on a too small local window.
Hence, a $6{\times}$ larger window is considered for the computation of $\rho^{\rm max}$, {\em i.e.} $6\sigma{\times}6\sigma$,
if $\sigma{\times}\sigma$ is the window size used for $\rho^{\sigma}_{\P\widehat{\M}_b}$.

Let us now focus on $\sigma$. 
In principle, 
forcing correlation between the PAN and any super-resolved band 
would be in contrast with the goal of preserving the consistency of the latter with its low-resolution version.
In practice, 
if the correlation is computed locally to a $R{\times}R=6{\times}6$ window, 
corresponding to the size of a low-resolution MS pixel mapped into the high-resolution space (PAN),
the related conditioning will interest only the high-frequency content which is missing in the low-resolution MS bands.
Beside this theoretical reasoning supporting a value $\sigma=R=6$,
we have also run an {\em ad hoc} experiment to analyze different settings.
In particular, 
Fig.~\ref{fig:sigma} shows the pansharpening results (zoomed crops were selected for an easier inspection) 
on a single sample band for $\sigma = 6,12, 24, 48$. 
Below each result, the spectral loss is also reported.
As it can be seen, the use of larger values of $\sigma$ deteriorates both the spectral (see numbers) 
and the spatial (see images) quality: larger $\sigma$s give rise to blurring phenomena.
On the other hand,
it has also to be observed that the use too small $\sigma$ values could be unsuited for the computation 
of the statistics involved in the correlation coefficient.
Eventually, in light of all the above considerations, we have fixed $\sigma=R=6$.
For additional details about the spatial loss, readers can refer to \cite{Ciot2022}.

\renewcommand{\image}[2]{\includegraphics[width=0.18\linewidth]{./figures/test_sigma/FR2_#1_crop#2.png}}
\begin{figure*}[ht]
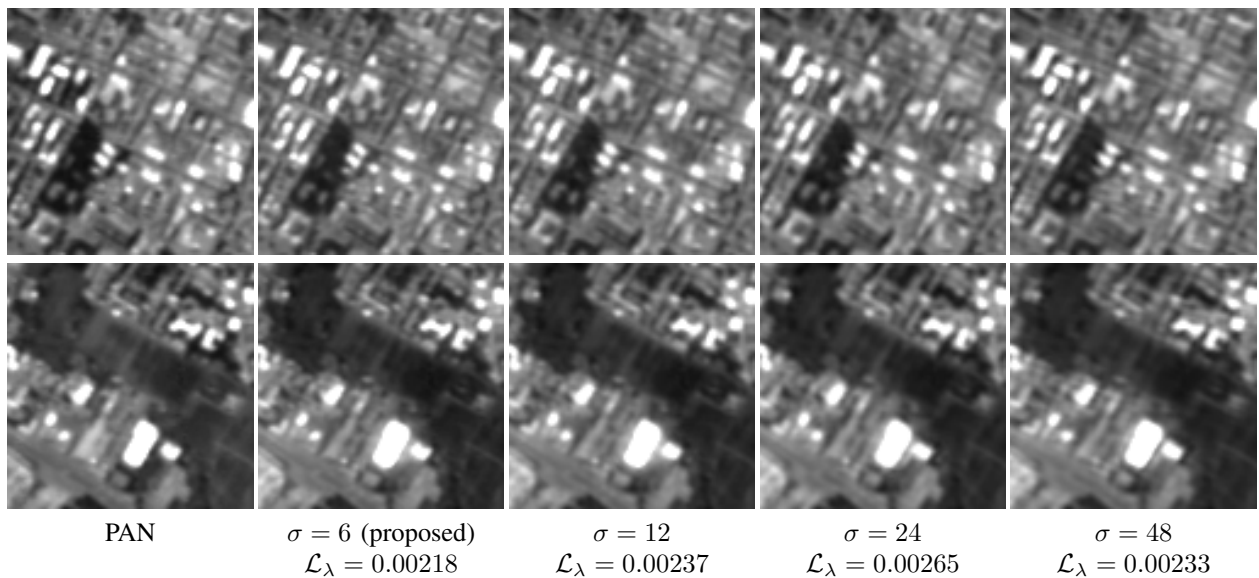

    \centering
    \setlength\tabcolsep{1pt}
    \begin{tabular}{ccccc}
        \image{PAN}{1}  & \image{S6}{1} & \image{S12}{1} & \image{S24}{1} & \image{S48}{1} \\ 
        \image{PAN}{2}  & \image{S6}{2} & \image{S12}{2} & \image{S24}{2} & \image{S48}{2} \\ 
        PAN & $\sigma=6$ (proposed) & $\sigma=12$& $\sigma=24$ & $\sigma=48$ \\
        & $\LL_\lambda = 0.00218$ & $\LL_\lambda = 0.00237$ & $\LL_\lambda = 0.00265$ & $\LL_\lambda = 0.00233$ 
    \end{tabular}    
    \caption{Impact of the scale parameter $\sigma$ (correlation scale) on spectral and spatial quality.
    From left to right: image crops (PAN) followed by the corresponding pansharpening results for a sample band, using increasing scale values ($\sigma = 6, 12, 24, 48$).
    The two crops come from a larger tile on which the pansharpening is run.
    Below each result, the spectral loss is reported obtained using 1000 tuning iterations.}
    \label{fig:sigma}
\end{figure*}

\subsection{Network configuration}
\label{sec:ablation}

\begin{table}
	\centering
	\scriptsize
	\setlength{\tabcolsep}{3pt}
    \begin{tabular}{ccccccc}
        \hline
        \# Layer & Residual & Kernels & Time & $D_\lambda$ ($\downarrow$) & $D_S$ ($\downarrow$) & $Q^\ast$ ($\uparrow$)\\
        \hline
        3 & Yes & (48,32,1) & 51' & 0.0312 & 0.0136 & \bf 0.9556\\
        \hline
        3 & No &  (48,32,1) & 51' & 0.0419 & 0.0119 & 0.9466\\
        3 & Yes & (32,16,1) & \bf 50' & 0.0333 & 0.0159 & 0.9513\\
        3 & No &  (32,16,1) & \bf 50' & 0.0489 & \bf 0.0084 & 0.9431\\
		3 & Yes & (64,32,1) & 61' & \bf 0.0296 & 0.0157 & 0.9552\\	
        3 & No & (64,32,1) & 61' & 0.0394 & 0.0096 & 0.9514\\
        \hline
        4 & Yes & (48,32,16,1) & 59' & 0.0311 & 0.0139 & 0.9553\\
        4 & No & (48,32,16,1) &	59' & 0.0449 & 0.0102 & 0.9454\\
        5 & Yes & (64,48,32,16,1) & 74' & 0.0305 & 0.0167 & 0.9533\\
        5 & No & (64,48,32,16,1)& 74' & 0.0395 & 0.0098 & 0.9511\\
        \hline
    \end{tabular}    
    \label{tab:ablation}
    \vspace{2mm}
    \caption{Network ablation results on dataset FR0. The proposed solution is on top.}
\end{table}

To validate the network architecture, we carried out experiments on the validation dataset FR0
aimed to assess the impact of network depth, width, and the use of a residual skip connection.
The compared solutions are summarized in Tab.~\ref{tab:ablation}, 
together with the corresponding scores and execution time.
At a first glance,
it can be observed that the proposed configuration (top row) achieves the best score on $Q^\ast$ balancing spectral and spatial accuracies.
Besides, the execution time is nearly the same as the lighter configuration (32,16,1).
A deeper analysis of these results reveals that deeper architectures tend to slightly favor the spectral consistency ($D_\lambda$) with respect to the 
spatial index ($D_S$).
The quantitative analysis also shows that the residual skip connections help to preserve spectral consistency keeping lower the $D_\lambda$ score.
Overall, the topological configuration (i.e., with or without residual skip connection) has a more relevant impact on the network behavior than its depth and width.

\subsection{Pretraining details}
\label{sec:pretrain}
At test time, the proposed network starts the tuning on the first band using pretrained initial weights.
These have been determined by a pretraining using the first band of the FR0 dataset.
The band and the corresponding PAN have been tiled in 100 patches of (PAN) size 240$\times$240, splitting in training (90) and validation (10).
The optimization has been carried out using ADAM in default configuration and with a learning rate of 1e-5,
run for 200 epochs using minibatches of 4 patches.
The same pretrained weights have been used on both reduced- and full-resolution images at test time.

\section{Comparative results and discussion}
\label{sec:results}
The experimental analysis of the proposed solution ends 
with the presentation of the numerical and visual comparative results obtained on the test datasets 
(summarized in Tab.~\ref{tab:datasets}, shown in Fig.~\ref{fig:datasets})
taken from the HS pansharpening challenge \cite{Vivo23}.
All numerical results, 
obtained on both reduced- and full-resolution datasets, 
are gathered in Tab.~\ref{tab:results}.
On the one hand, 
the pansharpening results obtained on reduced-resolution datasets are quantitatively compared 
in terms of $Q2^n$, SAM, ERGAS and PSNR.
On the other hand,
those obtained on full-resolution datasets are assessed through the spectral and spatial quality indexes $D_\lambda$ and $D_S$, respectively,
and their combination $Q^\ast$ (details are given in Sec.~\ref{sec:data}). 
Besides,
an overall accuracy index (OA) is reported in the last column of Tab.~\ref{tab:results}.
OA is computed as average, over all the four datasets, of $Q2^n$ or $Q^\ast$, 
using the one that applies (RR or FR, respectively).
OA is therefore a robust indicator of the consistency of the methods, 
accounting for their behavior at both resolutions,
that has been chosen as ultimate score in the challenge. 
The table gathers the comparative methods in three sets: challenge baselines (CS and MRA), challenge participants (teams), 
and recent deep learning solutions. 
Moreover, in the top row, we report the scores of the simple interpolator EXP, which serves as reference for the spectral accuracy at full-resolution, lacking ground-truths. Bold and underlined numbers highlight top and second best scores, respectively, excluding EXP, which does not perform any sharpening.

\newcommand{\D}{\small}

\begin{table*}[htbp]
    \centering
    \resizebox{0.95\textwidth}{!}{
    \begin{tabular}{|l|cccc|cccc|ccc|ccc|c|}
    \hline
       & \multicolumn{4}{c}{RR1} \vline & \multicolumn{4}{c}{RR2} \vline & \multicolumn{3}{c}{FR1} \vline  & \multicolumn{3}{c}{FR2} \vline & \multirow{2}{*}{OA}\\
       & $Q2^n$ & SAM & ERGAS & PSNR & $Q2^n$ & SAM & ERGAS & PSNR & $D_\lambda$ & $D_S$ & $Q^\ast$ & $D_\lambda$ & $D_S$ & $Q^\ast$ & \\
      \hline
      EXP                         & 0.5740 & 3.3203 & 3.6816 & 32.1257 & 0.5717 & 6.1252 & 7.0178 & 28.1808 & 0.0174 & 0.2889 & 0.6987 & 0.0159 & 0.2415 & 0.7464 & 0.6477\\ \hline
      \cite{Laben2000} GS         & 0.5080 & 14.104 & 7.9535 & 24.8517  & 0.7221 & 5.5997 & 5.9123 & 28.8292 & 0.2648  & \underline{0.0113} & 0.7269 & 0.2073 & 0.0344 & 0.7654 & 0.6806\\
      \cite{Aiazzi2007} GSA       & \underline{0.6912} & \underline{3.3297} & \underline{2.8750} & \underline{33.7928} & 0.8251 & 4.8277 & \underline{4.4092} & 29.7096 & 0.0512 & 0.0121 & \underline{0.9373} & 0.0555 & \underline{0.0033} & 0.9414  & \underline{0.8488} \\
      \cite{Vivone2015} AWLP      & 0.6166 & 4.8929 & 4.1890 & 28.7526  & 0.6935 & 7.6157 & 5.9395 & 25.4248 & \underline{0.0274} & 0.0490 & 0.9249 & \underline{0.0257} & 0.0367 & 0.9385 & 0.7934\\
      \cite{Aiazzi2006} MTF-GLP   & 0.6207 & 3.6398 & 3.9223 & 30.4626 & 0.7963 & 5.3705 & 4.7784 & 27.9672 & 0.0396 & 0.0341 & 0.9277 & 0.0331 & 0.0254 & 0.9423 & 0.8218\\
      \cite{Restaino2016} MF      & 0.6189 & 5.5063 & 7.2229 & 26.8007 & 0.8057 & 5.2975 & 4.7005 & 29.0872 & 0.0939 & 0.0663 & 0.8460 & 0.1044 & 0.0447 & 0.8556 & 0.7816\\
      \hline
      Team 1                      & 0.1286 & 10.553 & 41.028 & 12.0480 & 0.1280 & 14.015 & 81.360 & 11.2153 & 0.6331 & \bf 0.0008 & 0.3666 & 0.4900 & \bf 0.0006 & 0.5097 & 0.2832\\
      Team 2                      & 0.5263 & 3.7735 & 5.1628 & 28.8121 & 0.7595 & 5.3577 & 5.1664 & 28.9139 & 0.0831 & 0.0134 & 0.9046 & 0.1055 & 0.0108 & 0.8849 & 0.7688\\
      Team 3                      & 0.6849 & 4.1293 & 3.3394 & 32.6191 & 0.8137 & 5.2395 & 5.2888 & 29.4204 & 0.0837 & 0.0360 & 0.8833 & 0.0876 & 0.0131 & 0.9004 & 0.8206\\
      Team 4                      & 0.6402 & 5.0922 & 3.9272 & 30.6497 & 0.7341 & 6.7361 & 5.0231 & 27.2483 & 0.0338 & 0.0269 & \bf 0.9402 & 0.0336 & 0.0168 & \underline{0.9501} & 0.8162\\
      \hline
\cite{He19} HyperPNN1 & 0.6491 & 5.2565 & 3.2416 & 32.9277 & 0.8265 & 5.1380 & 6.0469 & 30.0159 & 0.1203 & 0.0117 & 0.8695 & 0.1198 & 0.0152 & 0.8668 & 0.8030\\
\cite{He19} HyperPNN2 & 0.6629 & 5.6340 & 3.2195 & 32.9648 & \underline{0.8348} & 5.2261 & 5.4397 & \underline{30.1385} & 0.1130 & 0.0116 & 0.8767 & 0.1064 & 0.0125 & 0.8824 & 0.8142\\
\cite{He20} HSpeNet1 & 0.6857 & 4.3354 & 2.8756 & 33.7916 & 0.8264 & 4.7937 & 5.5613 & 28.7095 & 0.0748 & 0.0401 & 0.8881 & 0.0714 & 0.0149 & 0.9148 & 0.8288\\
\cite{He20} HSpeNet2 & 0.6846 & 3.8857 & \bf 2.8272 & \bf 33.9206 & 0.8240 & \underline{4.2805} & 5.8318 & 27.4986 & 0.0709 & 0.0481 & 0.8845 & 0.0718 & 0.0212 & 0.9086 & 0.8254\\
      \hline
      R-PNN (proposed) 
      		& \bf 0.7196 & \bf 3.1713 & 2.9199 & 33.7633 & \bf 0.8353 &  \bf 4.2137 &  \bf 4.3426 & \bf 30.3403 & \bf 0.0214 & 0.0444 & 0.9352 & \bf 0.0226 & 0.0232 & \bf 0.9547 & \bf 0.8612\\
    \hline
    \end{tabular}
    }
    \vspace{1mm}
    \caption{Comparative numerical results.}
    \label{tab:results}
\end{table*}

Moving to the full-resolution datasets,
the analysis of the results becomes lesser linear.
On the spectral side ($D_\lambda$),
the proposed method ranks always first (the interpolator EXP is excluded as it does not perform any pansharpening as testifyed by the high spatial distortion $D_S$).
Instead, 
on the spatial consistency side
the proposed solution shows some limits.
Indeed, 
several competitors with good (small) values of $D_S$ 
show an exceeding spectral distortion $D_\lambda$
that knocks them out.
$Q^\ast$ balances the two distortion indexes providing a more reliable numerical figure.
According to $Q^\ast$, the proposed method ranks first on FR2
and third on FR1, though not far from the best (Team4).
This last one, however, does not seem to provide consistent results across the two scales
showing much worse results on the RR datasets.
Overall,
our method achieves the best OA with a consistent gap on GSA that ranks second.
As a final remark, it is worth to notice that, at both reduced- and full-resolutions,
the proposed method consistently outperforms the competitors from the spectral point of view.
In fact, it is always on top in terms of SAM, which is probably the most sensible index to the spectral signature,
and of $D_\lambda$ (considering the EXP as spectral reference at full-resolution).
This gives us an encouraging experimental response to a basic question about the opportunity to 
carry out band-wise pansharpening, as we do, ignoring the pixel spectral signature as a whole feature to account for in the fusion process.
Our experiments show that, despite a marginal optimization scheme has been used for computational reasons,
the pixels spectral signatures are very well preserved.

\newcommand{\clipwid}{0.105\linewidth}
\newcommand{\figScale}{1.35}
\newcommand{\imagerr}[1]{\includegraphics[width=\clipwid]{./figures/R_PNN_results/RR1/#1.png}}

\begin{figure*}[!htbp]
    \setlength\tabcolsep{1pt}
        \centering
    \begin{tabular}{cccccccc}    
        \imagerr{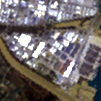} & \imagerr{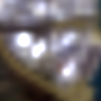} & \imagerr{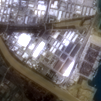} & \imagerr{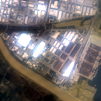} & \imagerr{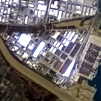} & \imagerr{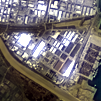} & \imagerr{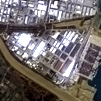}  &  \imagerr{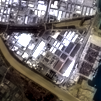} \\
        \D GT & \D EXP & \D R-PNN & \D GS & \D GSA & \D AWLP & \D MTF-GLP & \D MF \\[1mm]
        \imagerr{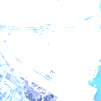} & \imagerr{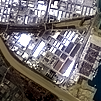} & \imagerr{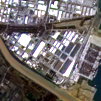} & \imagerr{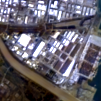} & \imagerr{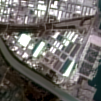} & \imagerr{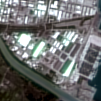} & \imagerr{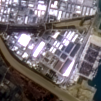}  &  \imagerr{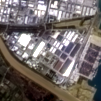} \\
        \D Team 1 & \D Team 2 & \D Team 3 & \D Team 4 & \D HyperPNN1 & \D HyperPNN2 & \D HSpeNet1 & \D HSpeNet2 \\     
 \multicolumn{8}{c}{(a)}\\[2mm] 
        \imagerr{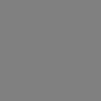} & \imagerr{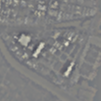} & \imagerr{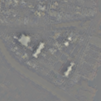} & \imagerr{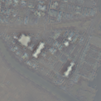} & \imagerr{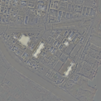} & \imagerr{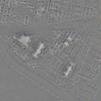} & \imagerr{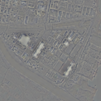}  & \imagerr{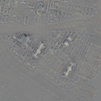} \\
		\D GT & \D EXP & \D R-PNN & \D GS & \D GSA & \D AWLP & \D MTF-GLP & \D MF \\[1mm]
        \imagerr{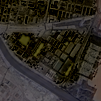} & \imagerr{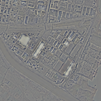} & \imagerr{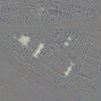} & \imagerr{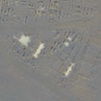} & \imagerr{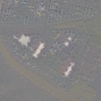} & \imagerr{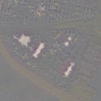} & \imagerr{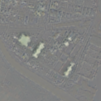}  & \imagerr{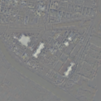} \\ 
        \D Team 1 & \D Team 2 & \D Team 3 & \D Team 4 & \D HyperPNN1 & \D HyperPNN2 & \D HSpeNet1 & \D HSpeNet2 \\ 
 \multicolumn{8}{c}{(b)}\\
\end{tabular}
\caption{Pansharpening results on RR1 (zoomed detail, see Fig.~\ref{fig:datasets}).
In (a) are shown the target GT and all compared solutions on three bands sampled in the visible spectrum
Wavelengths [nm]: 660 (red channel), 588 (green), 442 (blue).
In (b) are the corresponding error maps.}
\label{fig:RR1rgb}
\end{figure*}

\begin{figure*}[!htbp]
  \setlength\tabcolsep{1pt}
      \centering
    \begin{tabular}{cccccccc}
	\imagerr{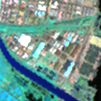} & \imagerr{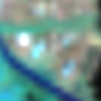} & \imagerr{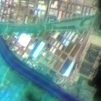} & \imagerr{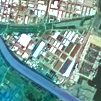} & \imagerr{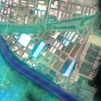} & \imagerr{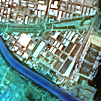} & \imagerr{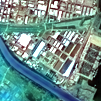}  &  \imagerr{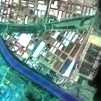} \\
        \D GT & \D EXP & \D R-PNN & \D GS & \D GSA & \D AWLP & \D MTF-GLP & \D MF \\[1mm]
        \imagerr{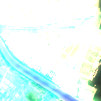} & \imagerr{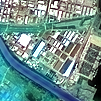} & \imagerr{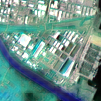} & \imagerr{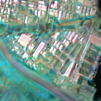} & \imagerr{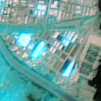} & \imagerr{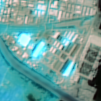} & \imagerr{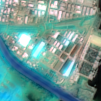}  &  \imagerr{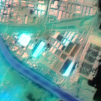} \\
        \D Team 1 & \D Team 2 & \D Team 3 & \D Team 4 & \D HyperPNN1 & \D HyperPNN2 & \D HSpeNet1 & \D HSpeNet2 \\    
 \multicolumn{8}{c}{(a)}\\[2mm] 
	\imagerr{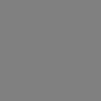} & \imagerr{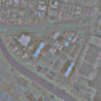} & \imagerr{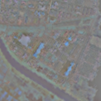}  & \imagerr{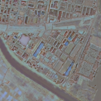} & \imagerr{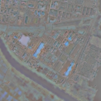} & \imagerr{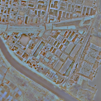} & \imagerr{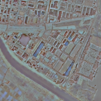}  & \imagerr{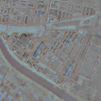}\\
        \D GT & \D EXP & \D R-PNN & \D GS & \D GSA & \D AWLP & \D MTF-GLP & \D MF \\ [1mm]
        \imagerr{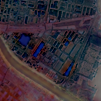} & \imagerr{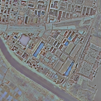} & \imagerr{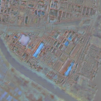} & \imagerr{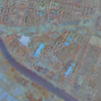} & \imagerr{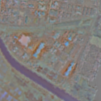} & \imagerr{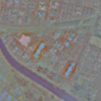} & \imagerr{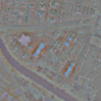}  & \imagerr{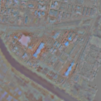} \\
        \D Team 1 & \D Team 2 & \D Team 3 & \D Team 4 & \D HyperPNN1 & \D HyperPNN2 & \D HSpeNet1 & \D HSpeNet2 \\
 \multicolumn{8}{c}{(b)}\\
\end{tabular}
\caption{Pansharpening results on RR1 (zoomed detail, see Fig.~\ref{fig:datasets}).
In (a) are shown the target GT and all compared solutions on three bands sampled outside the visible spectrum
Wavelengths [nm]: 2053 (red channel), 1229 (green), 770 (blue).
In (b) are the corresponding error maps.}
\label{fig:RR1fc}
\end{figure*}

\renewcommand{\imagerr}[1]{\includegraphics[width=\clipwid]{./figures/R_PNN_results/RR2/#1.png}}

\begin{figure*}[!htbp]
    \setlength\tabcolsep{1pt}
    \centering
    \begin{tabular}{cccccccc}    
        \imagerr{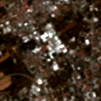} & \imagerr{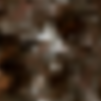} & \imagerr{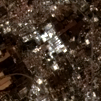} & \imagerr{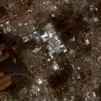} & \imagerr{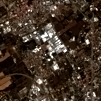} & \imagerr{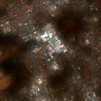} & \imagerr{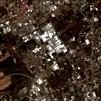}  &  \imagerr{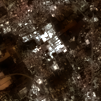} \\
        \D GT & \D EXP & \D R-PNN & \D GS & \D GSA & \D AWLP & \D MTF-GLP & \D MF \\[1mm]
        \imagerr{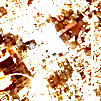} & \imagerr{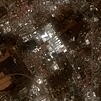} & \imagerr{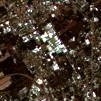} & \imagerr{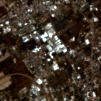} & \imagerr{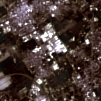} & \imagerr{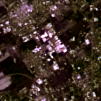} & \imagerr{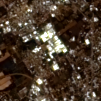}  &  \imagerr{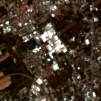} \\
        \D Team 1 & \D Team 2 & \D Team 3 & \D Team 4 & \D HyperPNN1 & \D HyperPNN2 & \D HSpeNet1 & \D HSpeNet2 \\     
 \multicolumn{8}{c}{(a)}\\[2mm] 
        \imagerr{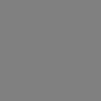} & \imagerr{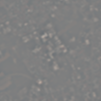} & \imagerr{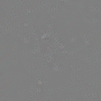} & \imagerr{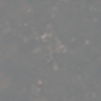} & \imagerr{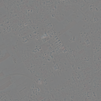} & \imagerr{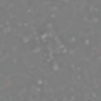} & \imagerr{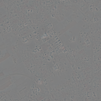}  & \imagerr{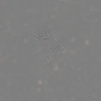} \\
		\D GT & \D EXP & \D R-PNN & \D GS & \D GSA & \D AWLP & \D MTF-GLP & \D MF \\[1mm]
        \imagerr{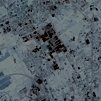} & \imagerr{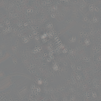} & \imagerr{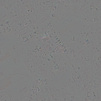} & \imagerr{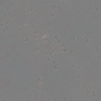} & \imagerr{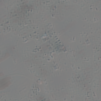} & \imagerr{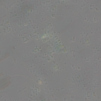} & \imagerr{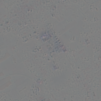}  & \imagerr{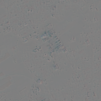} \\ 
        \D Team 1 & \D Team 2 & \D Team 3 & \D Team 4 & \D HyperPNN1 & \D HyperPNN2 & \D HSpeNet1 & \D HSpeNet2 \\ 
 \multicolumn{8}{c}{(b)}\\
\end{tabular}
\caption{Pansharpening results on RR2 (zoomed detail, see Fig.~\ref{fig:datasets}).
In (a) are shown the target GT and all compared solutions on three bands sampled in the visible spectrum
Wavelengths [nm]: 632 (red channel), 500 (green), 434 (blue).
In (b) are the corresponding error maps.}
\label{fig:RR2rgb}
\end{figure*}

\begin{figure*}[!htbp]
  \setlength\tabcolsep{1pt}
    \centering
    \begin{tabular}{cccccccc}
	\imagerr{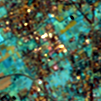} & \imagerr{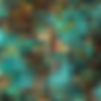} & \imagerr{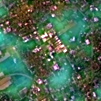} & \imagerr{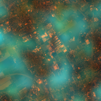} & \imagerr{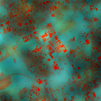} & \imagerr{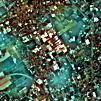} & \imagerr{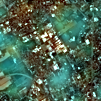}  &  \imagerr{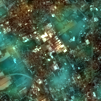} \\
        \D GT & \D EXP & \D R-PNN & \D GS & \D GSA & \D AWLP & \D MTF-GLP & \D MF \\[1mm]
        \imagerr{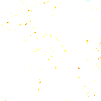} & \imagerr{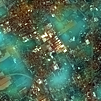} & \imagerr{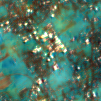} & \imagerr{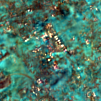} & \imagerr{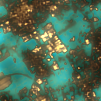} & \imagerr{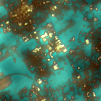} & \imagerr{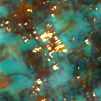}  &  \imagerr{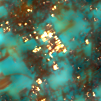} \\
        \D Team 1 & \D Team 2 & \D Team 3 & \D Team 4 & \D HyperPNN1 & \D HyperPNN2 & \D HSpeNet1 & \D HSpeNet2 \\    
 \multicolumn{8}{c}{(a)}\\[2mm] 
	\imagerr{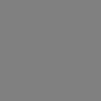} & \imagerr{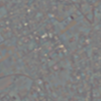} & \imagerr{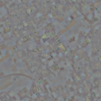}  & \imagerr{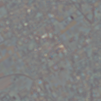} & \imagerr{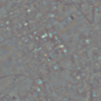} & \imagerr{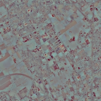} & \imagerr{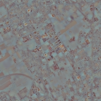}  & \imagerr{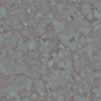}\\
        \D GT & \D EXP & \D R-PNN & \D GS & \D GSA & \D AWLP & \D MTF-GLP & \D MF \\ [1mm]
        \imagerr{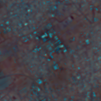} & \imagerr{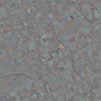} & \imagerr{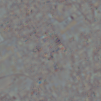} & \imagerr{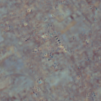} & \imagerr{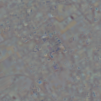} & \imagerr{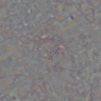} & \imagerr{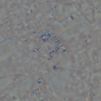}  & \imagerr{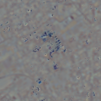} \\
        \D Team 1 & \D Team 2 & \D Team 3 & \D Team 4 & \D HyperPNN1 & \D HyperPNN2 & \D HSpeNet1 & \D HSpeNet2 \\
 \multicolumn{8}{c}{(b)}\\
\end{tabular}
\caption{Pansharpening results on RR2 (zoomed detail, see Fig.~\ref{fig:datasets}).
In (a) are shown the target GT and all compared solutions on three bands sampled outside the visible spectrum
Wavelengths [nm]: 1726 (red channel), 1251 (green), 750 (blue).
In (b) are the corresponding error maps.}
\label{fig:RR2fc}
\end{figure*}

\newcommand{\imager}[1]{\includegraphics[width=\clipwid]{./figures/R_PNN_results/FR1/#1.png}}
\newcommand{\imagefr}[1]{\includegraphics[width=\clipwid]{./figures/R_PNN_results/FR2/#1.png}}

\begin{figure*}[!htbp]
    \setlength\tabcolsep{1pt}
    \centering
    \begin{tabular}{cccccccc}
        \imager{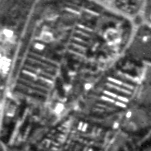} & \imager{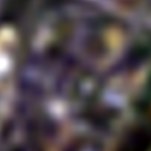} & \imager{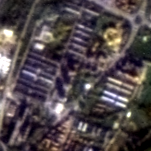} & \imager{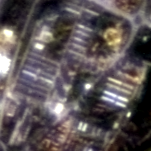} & \imager{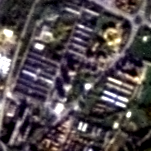} & \imager{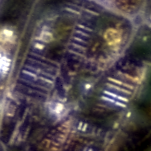} & \imager{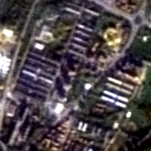}  &  \imager{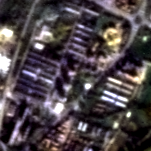} \\
        \D PAN & \D EXP & \D R-PNN & \D GS & \D GSA & \D AWLP & \D MTF-GLP & \D MF \\[1mm]
        \imager{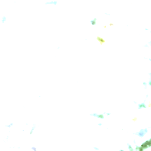} & \imager{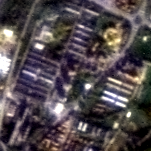} & \imager{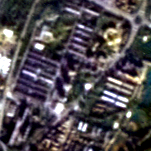} & \imager{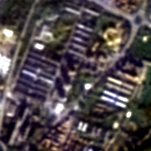} & \imager{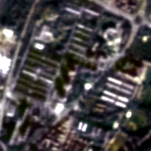} & \imager{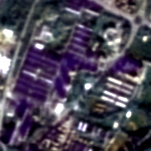} & \imager{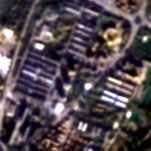}  &  \imager{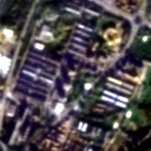} \\
        \D Team 1 & \D Team 2 & \D Team 3 & \D Team 4 & \D HyperPNN1 & \D HyperPNN2 & \D HSpeNet1 & \D HSpeNet2 \\     
 \multicolumn{8}{c}{(a)}\\[2mm] 
\imagefr{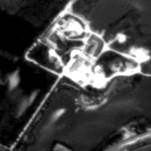} & \imagefr{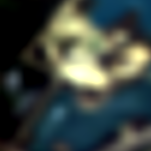} & \imagefr{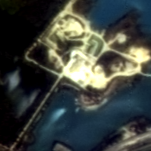} & \imagefr{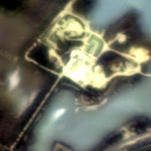} & \imagefr{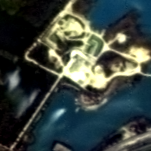} & \imagefr{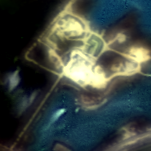} & \imagefr{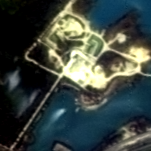}  &  \imagefr{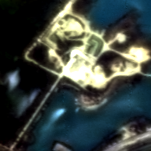} \\
        \D PAN & \D EXP & \D R-PNN & \D GS & \D GSA & \D AWLP & \D MTF-GLP & \D MF \\[1mm]
        \imagefr{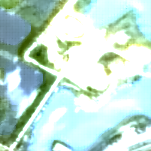} & \imagefr{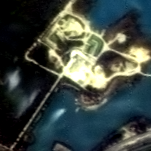} & \imagefr{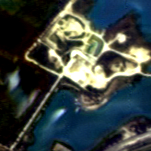} & \imagefr{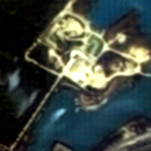} & \imagefr{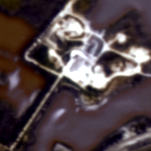} & \imagefr{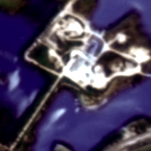} & \imagefr{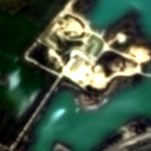}  &  \imagefr{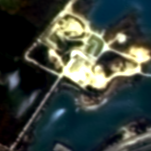} \\
        \D Team 1 & \D Team 2 & \D Team 3 & \D Team 4 & \D HyperPNN1 & \D HyperPNN2 & \D HSpeNet1 & \D HSpeNet2 \\    
 \multicolumn{8}{c}{(b)}\\
 \end{tabular}
\caption{Pansharpening results on FR1 (a) and FR2 (b) on three bands sampled in the visible spectrum;
PAN image followed by EXP and all compared methods.
Details on crop selection and sampled wavelengths for display are in Fig.~\ref{fig:datasets}.}
\label{fig:FR_RGB}
\end{figure*}

\begin{figure*}[!htbp]
    \setlength\tabcolsep{1pt}
    \centering
    \begin{tabular}{cccccccc}
        \imager{FR1_PAN_Z} & \imager{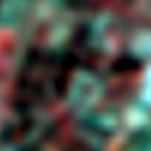} & \imager{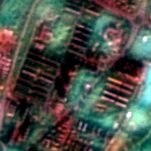} & \imager{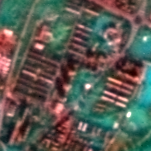} & \imager{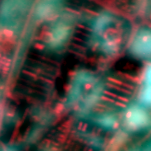} & \imager{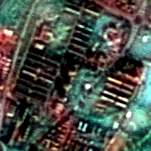} & \imager{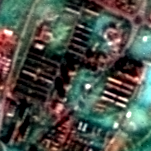}  &  \imager{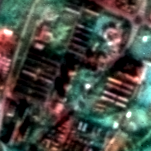} \\
        \D PAN & \D EXP & \D R-PNN & \D GS & \D GSA & \D AWLP & \D MTF-GLP & \D MF \\[1mm]
        \imager{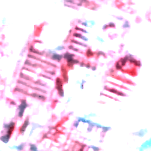} & \imager{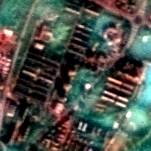} & \imager{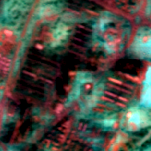} & \imager{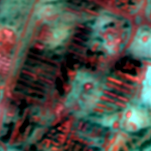} & \imager{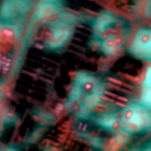} & \imager{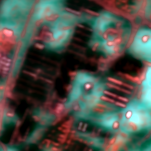} & \imager{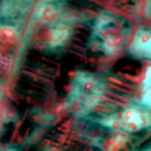}  &  \imager{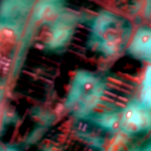} \\
        \D Team 1 & \D Team 2 & \D Team 3 & \D Team 4 & \D HyperPNN1 & \D HyperPNN2 & \D HSpeNet1 & \D HSpeNet2 \\     
 \multicolumn{8}{c}{(a)}\\[2mm] 
\imagefr{FR2_PAN_Z} & \imagefr{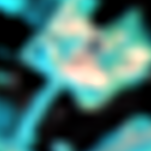} & \imagefr{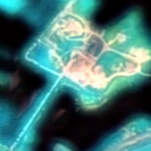} & \imagefr{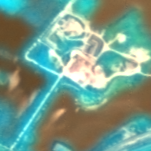} & \imagefr{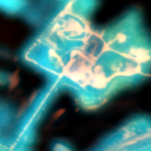} & \imagefr{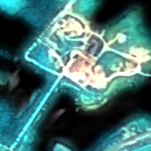} & \imagefr{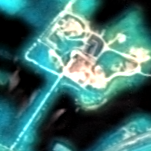}  &  \imagefr{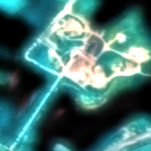} \\
        \D PAN & \D EXP & \D R-PNN & \D GS & \D GSA & \D AWLP & \D MTF-GLP & \D MF \\[1mm]
        \imagefr{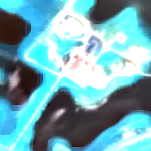} & \imagefr{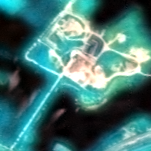} & \imagefr{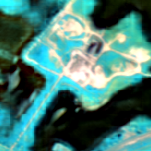} & \imagefr{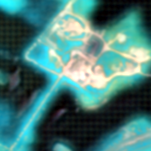} & \imagefr{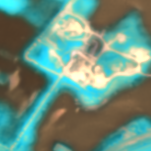} & \imagefr{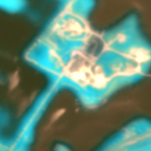} & \imagefr{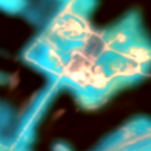}  &  \imagefr{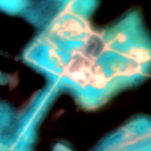} \\
        \D Team 1 & \D Team 2 & \D Team 3 & \D Team 4 & \D HyperPNN1 & \D HyperPNN2 & \D HSpeNet1 & \D HSpeNet2 \\    
 \multicolumn{8}{c}{(b)}\\
 \end{tabular}
\caption{Pansharpening results on FR1 (a) and FR2 (b) on three bands sampled outside the visible spectrum;
PAN image followed by EXP and all compared methods.
Details on crop selection and sampled wavelengths for display are in Fig.~\ref{fig:datasets}.}
\label{fig:FR_FC}
\end{figure*}

A careful inspection of these numerical results reveals a surprising behavior on reduced-resolution datasets 
by the proposed method which outperforms consistently all the compared solutions, on both datasets and with respect to all indicators,
with the exception dataset RR1, where on ERGAS and PSNR (recall these two metrics are highly correlated) HSpeNet2 performs slightly better.
Actually,
what is worth to remark is that while RR indexes are based on available ground-truths, the proposed solution is fully unsupervised (both in training and in tuning)
and, nonetheless, it does not seem to show any performance gap for this.

To conclude this experimental survey, 
we present some sample pansharpening results for both reduced- 
and full-resolution datasets in Fig.~\ref{fig:RR1rgb}-\ref{fig:RR2fc} and Fig.~\ref{fig:FR_RGB}-\ref{fig:FR_FC}, respectively.
For each dataset, 
only a representative zoomed detail (crops within yellow boxes in Fig.~\ref{fig:datasets}) is shown
and, for a more comprehensive analysis,
both RGB (Fig.~\ref{fig:RR1rgb}, \ref{fig:RR2rgb}, \ref{fig:FR_RGB}) and false-color (Fig.~\ref{fig:RR1fc}, \ref{fig:RR2fc}, \ref{fig:FR_FC}) bands subsets are displayed.
In fact, while RGB bands are all well correlated with the PAN,
other bands outside the visible spectrum are lesser correlated, 
hence more critical from the fusion perspective and worth to inspect.
For the reduced-resolution case,
in addition to the reference ground-truth (GT) and to the expanded version (EXP) of the input HS bands,
useful for a direct spectral comparison of the pansharpening results,
the error images are also shown.
These latter clearly show that the errors are much more severe (for both the datasets) outside the visible spectrum (see false-color),
with the occurrence for all the methods of both spectral and spatial distortion phenomena.
However, among all the compared methods,
the proposed seems to mitigate better than others both spectral and spatial distortions.
Some methods clearly fail, {\em e.g.} Team 1, and they are reported only for the sake of completeness.

Moving to full-resolution results,
the evaluation becomes even more difficult and subjective, lacking reference ground-truths.
In Fig.~\ref{fig:FR_RGB} are shown the pansharpening results obtained on both FR1 (a) and FR2 (b),
limited to some selected bands of the visible spectrum, roughly corresponding to the red, green and blue channels.  
Fig.~\ref{fig:FR_FC} gathers, instead, the same results limited to other bands outside the visible range in false colors.
In both cases, together with the PAN, that is the spatial reference, 
it is also shown the upscaled HS (EXP) that can serve as spectral reference 
for quality assessment.
Likewise the reduced-resolution case,
we displayed zoomed details of the full pansharpening obtained on the images shown in Fig.~\ref{fig:datasets}.
Focusing on the FR1 detail,
we can observe relatively good results provided by the proposed, GSA, MTF-GLP, MF, Teams 3 and 4, HSpeNet1 and HSpeNet2, especially in the RGB space.
False-color results (Fig.~\ref{fig:FR_FC} (top)), 
instead, highlight some problems occurring for spectral bands outside the visible range.
For example, GSA seems to be unable to sharpen the interested bands.
The same for Teams 3 and 4 and for the HSpeNet variants.
The differences among the methods are more evident for the clip of FR2 due to the presence of water.
In this case, the spectral distortions are quite severe in several cases, {\em e.g.} GS, HyperPNN1, HyperPNN2, HSpeNet1,
but, more interestingly,
there can be noticed some PAN patterns (on the water basin), not present in false-color bands (see EXP on the bottom line),
which are added to the pansharpened images. 
Of course,
aware of the subjectiveness of these last considerations,
we leave the final say to readers who can add their own perspective to our observations and numbers.
In this regard,
for the sake of fairness,
we have also to remind that the teams of the challenge had a limited time to validate their design choices,
which somehow explains some unsatisfactory results.
\section{Conclusions}
\label{sec:conclusions}
In this work we have presented a novel deep-learning-based method for HS pansharpening.
The proposed approach requires a baseline CNN model for single-band pansharpening trainable/tunable
in an unsupervised manner, without resolution downgrade.
To this aim, we resorted to a recently proposed 4/8-band pansharpening model \cite{Ciot2022}, 
suitably adapted to the single-band case.
The baseline model is sequentially used for a band-wise pansharpening 
where the current application leverages on the model parameters adjusted on the previous band,
running a few tuning iterations to let them fitting the current (target) band.
The tuning is feasible thanks to the use of an unsupervised loss and the number of iterations is related to 
the ``spectral'' distance between the target band and its preceding one from which the model is inherited.

The advantages of the proposed method are: 
(i) the method is fully-unsupervised 
and does not require training data other than the same target image; 
(ii) it can be applied to any PAN-HS dataset, 
with no need to have a prefixed number of HS bands; 
(iii) the learning process, which is interleaved with the band-wise inference steps,
does not require resolution downgrade, a common but limiting option in pansharpening;
(iv) the method ensures good generalization properties thanks to the target-adaptive tuning;
(v) considering that tuning iterations are involved, the computational complexity is relatively limited 
especially if the baseline CNN is a lightweight network as actually is.

Despite the very good results achieved by the proposed approach,
there is still room left for improvement. 
More specifically, special attention should be put on the spatial consistency loss term used for the spectral bands that have no overlap with the PAN bandwidth.
Actually, this falls in the more general problem of spatial quality assessment of pansharpened images, which is well known to be far to be solved
\cite{Vivo21, Arie22, Scar2022}, becoming even more challenging in the hyperspectral case.
Another point that is worth investigating is the model propagation rule
and the iteration amount per band having an impact on the computational load.

In order to ensure full reproducibility of our research outcomes the code is
made available at \git.

\section{Acknowledgment}
\label{sec:acknowledgment}

The authors would like to thank the “National Biodiversity Future Center” (identification code CN00000033, CUP B83C22002930006) on ‘Biodiversity’,  financed under the National Recovery and Resilience Plan (NRRP),  Mission 4,  Component 2,  Investment 1.4 “Strengthening of research structures and creation of R{\&}D ‘national champions’ on some Key Enabling Technologies” - Call for tender No. 3138 of 16 December 2021,  rectified by Decree n.3175 of 18 December 2021 of Italian Ministry of University and Research funded by the European Union – NextGenerationEU; Award Number: Project code CN{\_}00000033, Concession Decree No. 1034 of 17 June 2022 adopted by the Italian Ministry of University and Research,  CUP - CUP B83C22002930006 ,- Project title “National Biodiversity Future Center - NBFC” .

\bibliographystyle{IEEEtran}
\bibliography{IEEEabrv,Bibliography}
\end{document}